\documentclass[11pt,a4paper]{article}
\pdfoutput=1
\usepackage{amsfonts,amssymb,amsmath,epsfig, graphicx,yfonts,jheppub}

\newcommand{\be}{\begin{equation}}
\newcommand{\ee}{\end{equation}}
\newcommand{\bea}{\begin{eqnarray}}
\newcommand{\eea}{\end{eqnarray}}

\newcommand{\mt}[1]{\textrm{\tiny #1}}
\newcommand{\cf}{{\cal F}}
\newcommand{\cb}{{\cal B}}

\newcommand{\prt}{\partial}
\newcommand{\vev}[1]{\langle #1\rangle}
\def\nc {N_\mt{c}}
\def\nf {N_\mt{f}}
\def\uh {u_\mt{H}}
\def\gym {g_\mt{YM}}

\title{Thermodynamics of anisotropic branes}

\author[a]{Daniel \'Avila,}
\author[b]{Daniel Fern\'andez,}
\author[a]{Leonardo Pati\~no,}
\author[c]{and Diego Trancanelli} 
\affiliation[a]{Departamento de F\'isica, Facultad de Ciencias, Universidad Nacional Aut\'onoma de M\'exico, \\  A.P. 70-542, M\'exico D.F. 04510, Mexico} 
\affiliation[b]{Max-Planck-Institut f\"ur Physik, F\"ohringer Ring 6, 80805 M\"unchen, Germany} 
\affiliation[c]{Institute of Physics, University of S\~ao Paulo, 05314-970 S\~ao Paulo, Brazil} 

\abstract{
We study the thermodynamics of flavor D7-branes embedded in an anisotropic black brane solution of type IIB supergravity. The flavor branes undergo a phase transition between a `Minkowski embedding', in which they lie outside of the horizon, and a `black hole embedding', in which they fall into the horizon. This transition depends on the black hole temperature, its degree of anisotropy, and the mass of the flavor degrees of freedom. It happens either at a critical temperature or at a critical anisotropy. A general lesson we learn from this analysis is that the anisotropy, in this particular realization, induces similar effects as the temperature. In particular, increasing the anisotropy bends the branes more and more into the horizon. Moreover, we observe that the transition becomes smoother for higher anisotropies.
}
  

\emailAdd{davhdz06@ciencias.unam.mx} 
\emailAdd{danielf@mpp.mpg.de}
\emailAdd{leopj@ciencias.unam.mx} 
\emailAdd{dtrancan@if.usp.br} 

\begin{document}  

\hfill{MPP-2016-275}

\maketitle
\setlength{\parskip}{3pt}


\section{Introduction}

It is by now a well-established entry of the AdS/CFT dictionary \cite{duality} that (a small number of) flavor degrees of freedom in the boundary gauge theory can be described, at strong coupling, in terms of (probe) flavor branes embedded in the dual geometry \cite{flavor}. For the case of ${\cal N}=4$ super Yang-Mills (SYM) in four dimensions with gauge group $SU(\nc)$ and $\nf$ flavors, this is done by considering $\nf$  D7-branes (with $\nf\ll \nc$, to avoid back-reaction) oriented along the $AdS_5$ directions and wrapping a $S^3\subset S^5$. At finite temperatures, the D7-branes have been shown to display an interesting thermodynamic behavior \cite{thermobrane}, with a first order transition taking place between a phase in which they lie outside of the black hole horizon -- the so-called `Minkowski embedding' -- and a phase in which they fall into it -- the `black hole embedding'. In the gauge theory, this corresponds to a transition between a discrete, gapped meson spectrum and a continuous, gapless distribution of excitations. 

Extending the holographic study of fundamental matter and its transitions to theories other than ${\cal N}=4$ SYM is clearly interesting in its own right. In fact, this has already been realized in various settings, see \cite{extensions} for a few examples and \cite{reviewmesons} for a review.  A special motivation to look for more realistic theories comes from the realization that the quark-gluon plasma (QGP) formed in the collision of relativistic heavy ions \cite{rhic} behaves as a strongly coupled fluid \cite{fluid}. This has prompted in recent years the use of holography as a possible tool to describe such a system, in what has grown into a rather broad and active area of research. An overview of applications of holography to strongly coupled plasmas can be found, e.g., in \cite{reviewmateosetal}. 

In practice, the class of plasmas that can be described using holography is quite different from the QGP \cite{soup}. One can nonetheless try to incorporate features of the real-world plasma in holographic models, to gain a qualitative understanding of its strong coupling behavior, which is of otherwise difficult access through ordinary field theoretical techniques.  One such feature is the intrinsic anisotropy produced by having a privileged spatial direction in the problem, namely the direction of the beam of the scattering particles, called $z$-direction in the following. This anisotropy in the initial conditions of the collisions results into a so-called `elliptic flow' of the plasma and in a peculiar phenomenology \cite{olli}. 

In the context of ${\cal N}=4$ SYM, this anisotropic state can be modeled at strong coupling in a {\it top-down} string construction by the introduction of a non-trivial axion field in the geometric background, as we shall review below \cite{MT,MT2}. On the gauge theory side, one considers the action \cite{ALT}
\be
S=S_{{\cal N}=4}+\frac{1}{8\pi^2}\int \theta(z)\, \mathrm{Tr}F\wedge F\,,
\qquad 
\theta(z)\propto z \,,
\label{actionQFT}
\ee
which is a marginally relevant deformation of the ${\cal N}=4$ SYM action, where the proportionality constant between the theta-angle $\theta(z)$ and $z$ has dimensions of energy and will be related to the parameter $a$ that we shall introduce in the next section. The rotational $SO(3)$ symmetry in the space directions is broken by the new term down to $SO(2)$ in the $xy$-plane. 

Various observables have been computed for this plasma, see for example \cite{observables} for a non-exhaustive list.\footnote{More recently, interesting unstable phases associated to this particular implementation of the anisotropy have been discovered in \cite{instabilities}.} However, one of the important phenomena that has so far eluded scrutiny are precisely the phase transitions of fundamental matter mentioned above. The theory in (\ref{actionQFT}) has, in fact, only matter fields in the adjoint representation of the $SU(\nc)$ gauge group. $\nf$ flavors of scalars and fermions in the fundamental representation may nonetheless be introduced and, with an abuse of language, we will refer to these fundamental fields indistinctly as `quarks'. 

The main objective of this paper is to extend the study of \cite{thermobrane} to the thermal, anisotropic state of \cite{MT,MT2}, and to study how the anisotropy affects the physics of fundamental matter transitions. An important characteristic of the background is the presence of a conformal anomaly arising during renormalization, which introduces a length scale $\mu$ \cite{MT2}. As a consequence, the physics of the transitions will depend, separately, on the black hole temperature, its degree of anisotropy and the mass of the flavor degrees of freedom, all divided by $\mu$ in order to get dimensionless ratios. To simplify the analysis, we shall fix here the value of $\mu$ (in practice, by fixing the position of the black hole horizon) and consider instead two independent dimensionless ratios given by $a/M_\mt{q}$ and $T/M_\mt{q}$. We compute how the free energy depends on these two ratios for fixed $T/a$ and $a/T$, respectively. 
The punch line of our analysis will be that the anisotropy has very similar effects to those of the temperature: increasing the anisotropy bends the flavor branes more and more into the horizon facilitating the transition between Minkowski and black hole embeddings. We also discover a {\it critical anisotropy} at which, for fixed temperature and quark masses, the phase transition takes place. This is quite similar to what happens in the case of infinitely heavy quarkonia, whose anisotropy-induced melting was studied in \cite{melting}.

This paper is organized as follows. In Sec.~\ref{sec2} we review the gravity set-up and in Sec.~\ref{sec3} we introduce probe D7-branes, considering both their black hole embedding and Minkowski embedding. In Sec.~\ref{sec4} we renormalize the brane on-shell action and compute the corresponding free energies. We conclude in Sec.~\ref{concl} with a discussion of our results. Some of the more technical details of our computations are contained in a series of appendices. 


\section{Review of the gravity set-up}
\label{sec2}

To model the gauge theory (\ref{actionQFT}) at strong coupling and in a thermal state, we consider the type IIB supergravity geometry found in \cite{MT,MT2} and given by
\bea
ds^2=\frac{L^2}{u^2}\left(-{\cal B} {\cal F} \, dt^2+dx^2+dy^2+e^{-\phi} dz^2 +\frac{du^2}{{\cal F}}\right)+L^2\, e^{\frac{1}{2}\phi}d\Omega_5^2\,,
\label{metric}
\eea
with $\Omega_5$ the volume form of a round 5-sphere. The dilaton $\phi$ and the metric fields ${\cal B}$ and ${\cal F}$ are solely functions of the radial coordinate $u$. This geometry contains a black brane, whose horizon is found at $u=\uh$, where $\cf$ vanishes. The boundary of the space in these coordinates is at $u=0$. The gauge theory coordinates are $(t,x,y,z)$: we refer to the $z$-direction as the longitudinal (or anisotropic) direction and to $x$ and $y$ as the transverse directions. The radius $L$ is set to unity in the following, without loss of generality. The other supergravity fields that are turned on are the forms
\bea
F_{(5)} =4 \left(\Omega_5 + \star \Omega_5\right)\,,\qquad F_{(1)}=a \, dz
\eea
with $a$ being a parameter with dimensions of energy that controls the degree of anisotropy between the longitudinal $z$-direction and the orthogonal $xy$-plane. The potential for the 1-form is a linear axion, $\chi=a\, z$, with fixed radial profile, which is responsible for maintaining the system in an anisotropic equilibrium state.

The metric functions and the dilaton are known analytically in limiting regimes of low and high temperature \cite{MT2}, while for intermediate regimes one has to resort to numerics to solve Einstein field equations \cite{MT2}. For $u\to 0$ and any value of $a$, they asymptote to the $AdS_5\times S^5$ metric, $\cf=\cb=1$ and $\phi=0$, while for $a=0$ they reduce to the black D3-brane solution
\be
\cb=1\,, \qquad \phi=\chi=0\,, \qquad \cf=1-\frac{u^4}{\uh^4}\,,
\label{isometric}
\ee
with temperature and entropy density given by
\be
T_\mt{iso}=\frac{1}{\pi\uh}\,, \qquad s_\mt{iso}= \frac{\pi^2}{2} \nc^2 T^3 \,.
\label{siso}
\ee
The anisotropic geometry has temperature and entropy density given instead by \cite{MT}
\be
T=\frac{e^{-\frac{1}{2}\phi_\mt{H}}\sqrt{{\cal B}_\mt{H}}(16+a^2 \uh^2 e^{\frac{7}{2}\phi_\mt{H}})}{16 \pi \uh}\,,
\qquad
s=\frac{\nc^2}{2\pi\uh^3}e^{-\frac{5}{4}\phi_\mt{H}}\,,
\label{temperature}
\ee
where $\phi_\mt{H}\equiv\phi(u=\uh)$ and $\cb_\mt{H}\equiv\cb(u=\uh)$. This entropy density interpolates smoothly between the isotropic scaling (\ref{siso}) for small $a/T$ and the scaling \cite{MT,ALT}
\be
s \simeq 3.21\, \nc^2 T^3 \left(\frac{a}{T}\right)^\frac{1}{3} \,,
\label{saniso}
\ee
for large $a/T$, so that the solution can be thought of as a domain-wall interpolating between AdS in the UV and a Lifshitz scaling in the IR. More details can be found in \cite{MT,MT2}.


\section{Flavor D7-brane embeddings}
\label{sec3}

The scope of this paper is to study the thermodynamics of fundamental matter coupled to (\ref{actionQFT}), which can be done at strong coupling by adding flavor D7-branes \cite{flavor} to the geometry described above. The whole system can be thought of as a D3/D7/D7 construction with $N_\mt{D7}$ anisotropy-inducing D7-branes\footnote{These $N_\mt{D7}$ branes are smeared homogeneously along the $z$-direction and give rise to a density $n_\mt{D7}=N_\mt{D7}/L_z$ of extended charges, with $L_z$ being the (infinite) length of the $z$-direction. This charge density is related to the anisotropy parameter $a$ through $a=\gym^2 n_\mt{D7}/4\pi$ \cite{MT2}.}  and $N_\mt{f}$ flavor D7-branes oriented as follows:
\bea
\begin{array}{l| cccc|c|ccc}
& t & x & y & z & u & \vartheta & \varphi & \Omega_3 \\
\hline 
N_\mt{c} ~~ \mbox{  D3 } & \times & \times & \times & \times & & & & \\
N_\mt{D7} ~ \mbox{  D7 } &  \times &  \times & \times & & &  \times &  \times &  \times \\
N_\mt{f}~ ~~\mbox{ D7 } &  \times &  \times &  \times & \times & \times &  & & \times 
\end{array}\,,
\eea
where $(\vartheta,\varphi,\Omega_3)$ are the coordinates on the $S^5$. The number of anisotropy inducing branes, $N_\mt{D7}$, is arbitrary, and in fact can be as large as desired, resulting in a fully back-reacted geometry. On the other hand, the number of flavor branes, $N\mt{f}$, must be small when compared to $N_\mt{c}$, in order to avoid back-reaction and to consider them as probes. We emphasize that this construction is entirely top-down and the resulting geometry can be obtained from string theory.

As mentioned in the Introduction, the flavor D7-branes have two possible embeddings, one in which they lie outside of the horizon, called `Minkowski embedding', and one in which they extend inside of the horizon, called `black hole embedding'. We now proceed to obtain the radial profile of these branes in both embeddings. To do so, it is convenient to start from the string frame metric (\ref{metric}) and to change the radial coordinate from $u$ to $\rho(u)$ with
\be
\frac{d \rho}{d u} = \frac{-\rho}{u \sqrt{\cf(u) \, e^{\frac{1}{2}\phi(u)}}}\,.
\label{defro}
\ee
We set $\rho(\uh)=1$, resulting in
\be
\rho = \exp{\int_{\uh}^u \frac{-d u'}{u'\sqrt{\cf(u')\, e^{\frac{1}{2}\phi(u')}}}}\,,
\label{defro2}
\ee
which can be inverted to obtain $u(\rho)$, needed in the following. In this new coordinate, the metric takes the form
\be
d s^2 = \frac{1}{{u(\rho)}^2} \left( -\cf(\rho)\cb(\rho) d t^2 + d x^2 + d y^2 +e^{-\phi(\rho)}d z^2 \right) +
\frac{e^{\frac{1}{2}\phi(\rho)}}{\rho^2} \left( d \rho^2 + \rho^2 d \Omega{^2_5} \right)\,.
\ee
The boundary is found at $\rho \rightarrow \infty$. The metric on the $S^5$ can be explicitly written as
\be
d \Omega{^2_5} = d \vartheta^2 + \cos^2 \vartheta \, d \varphi^2+ \sin^2 \vartheta \, d \Omega{^2_3}\,.
\ee
As seen above, the D7-branes' world-volume wraps the $(t,x,y,z,\Omega_3)$ directions and it also extends along the radial direction $\rho$. To find their radial profile we need to compute the induced metric on the world-volume and solve the equations of motion for the embedding functions $\vartheta(\rho)$ and $\varphi(\rho)$. We analyze the black hole and Minkowski embeddings separately.


\subsection{Black hole embedding}

In the case of a black hole embedding, the radial profile is specified by $\psi(\rho) \equiv \cos \vartheta (\rho)$ and $\varphi=$ constant. The induced metric is given by
\be
{\left(d \rho^2 + \rho^2 d \Omega{^2_5} \right)}_\mt{ind} = \left( 1 + \frac{\rho^2 \psi'^2 }{1-\psi^2} \right) d\rho^2 +
\rho^2 \left( 1 - \psi^2 \right)  d \Omega{^2_3}\,,
\ee
where the prime denotes a derivative with respect to $\rho$. Including the other world-volume directions, this results in
\be
\sqrt{-g_\mt{ind}} \propto  \frac{e^{\frac{1}{2}\phi}}{\rho\, u^4} \sqrt{\cf \cb} \left( 1 - \psi^2 \right) \sqrt{ 1 - \psi^2 + \rho^2  \psi'^2}\,,
\ee
where $u$ and the fields $\phi$, ${\cal F}$, and ${\cal B}$ have, again, to be considered functions of $\rho$. Notice that there is no Wess-Zumino term in the action, due to the particular orientation of the branes. To determine the profile $\psi (\rho)$ we solve its equation of motion. This follows from the action $S \propto \int e^{-\phi} \sqrt{-g_\mt{ind}}$ and reads
\be
\prt_{\rho} \left(P(\rho) \frac{\rho^5 (1-\psi^2)  \psi'}{\sqrt{ 1 - \psi^2 + \rho^2  \psi'^2 }} \right) +
P(\rho) \rho^3 \frac{3\psi(1-\psi^2)+2\rho^2\psi \psi'^2}{\sqrt{1-\psi^2+\rho^2 \psi'^2}} = 0\,,
\label{eqbh}
\ee
where\footnote{In the isotropic case, this function would simply be given by ${P(\rho)}_\mt{iso} = 1 - 1/\rho^8.$}
\be
P(\rho) = e^{- \frac{1}{2} \phi} {\left( \frac{1}{\rho\, u} \right)}^4 \sqrt{\cf \cb} .
\label{defp}
\ee
This equation is highly non-linear and has to be solved numerically, as we are going to do in a moment. The near-horizon expansion of $\psi$ needed to this scope is reported in App.~\ref{AppA}. 


\subsection{Minkowski embedding}

In the case of a Minkowski embedding, the radial profile is conveniently specified by $\varphi=$ constant and a function $R(r)$, with $R=\rho \cos \vartheta$ and $r=\rho\sin \vartheta$. This results in the induced metric (the prime is now a derivative with respect to $r$)
\be
{\left( d  \rho^2 + \rho^2 d  \Omega{^2_5} \right)}_\mt{ind} = \left( 1 +  R'^2 \right) d  r^2 +
r^2  d \Omega{^2_3}
\ee
and the action
\be
\sqrt{-g_\mt{ind}} \propto  \frac{e^{\frac{1}{2}\phi}}{u^4 \rho^4} r^3 \sqrt{\cf \cb} \sqrt{1 +  R'^2}\,.
\ee
The equation of motion is found to be given by
\be
\frac{d  P (\rho)}{d  R} r^3 \sqrt{1 +  R'^2} = \prt_r \left( P(\rho) \frac{r^3  R'}{\sqrt{1 +  R'^2}}  \right)\,,
\ee
where $P(\rho)$ is the same as above. Since this is a function of $\rho = \sqrt{r^2 + R(r)^2}$, the derivatives with respect to $\rho$ and $r$ can be calculated in terms of $P'(\rho) = d P / d \rho$. One obtains
\be
P \left( \sqrt{r^2 + R^2} \right) \prt_r \left( \frac{r^3  R'}{\sqrt{1 +  R'^2}} \right) =
P'\left( \sqrt{r^2 + R^2} \right) r^3\frac{(1 +  R'^2) R - r R'}{\sqrt{1 +  R'^2} \sqrt{r^2 + R^2}} \,,
\label{eqm}
\ee
which, again, can be solved numerically, as we do next.


\subsection{Numerical results}

\begin{figure}[h!]
    \begin{center}
        \includegraphics[width=0.69\textwidth]{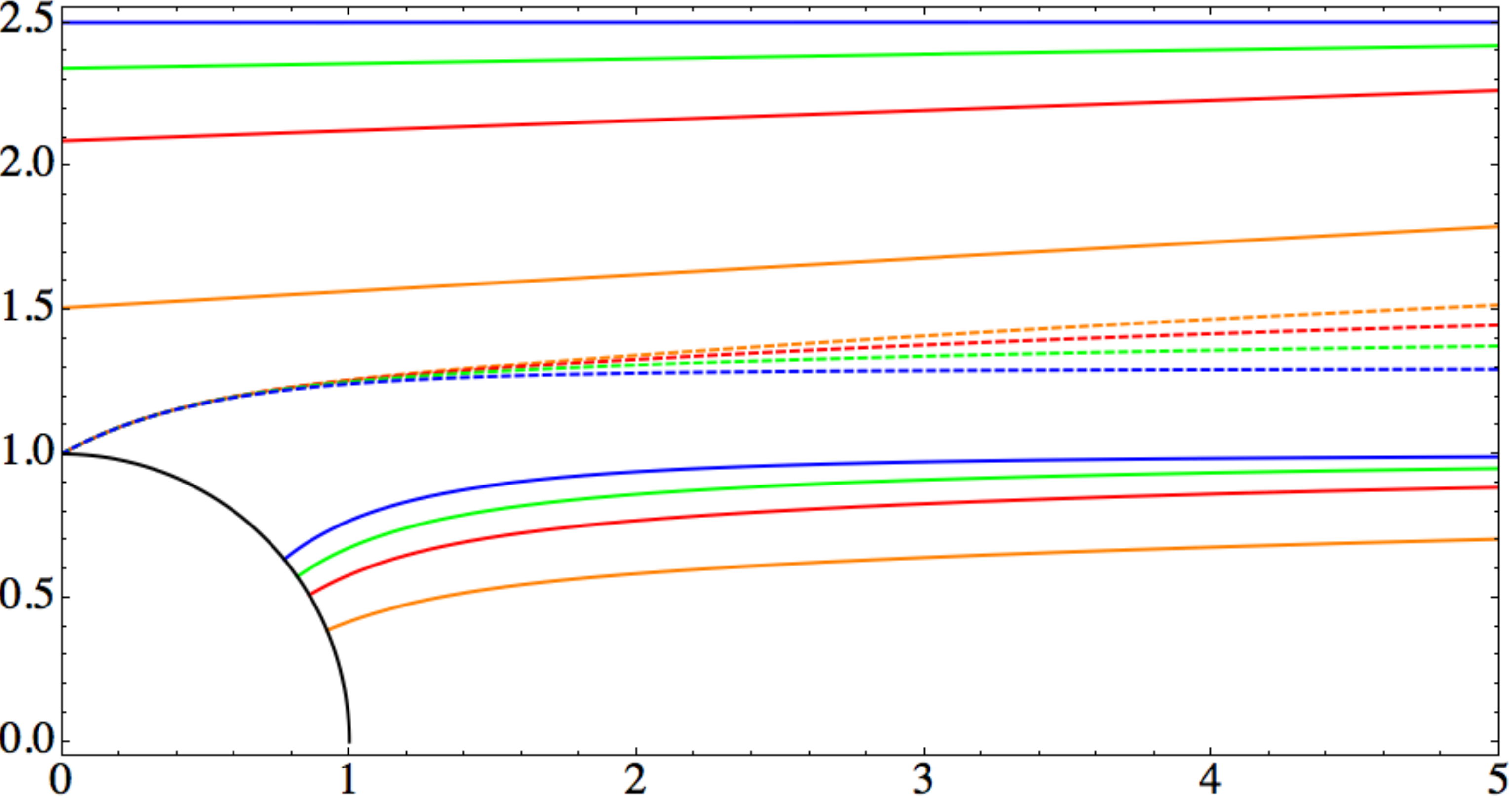}
        \put(-313,140){$R$}
        \put(-20,-5){$r$}
        \caption{Profiles for various D7-brane embeddings in the $(R,r)$-plane. The black circle represents the horizon at $\rho=1$. Blue, green, red, and orange continuous lines (top to bottom in upper and lower group of curves) correspond respectively to $a / T = \{0, 6.0, 13.7, 64.1 \}$. The four continuous curves on the top correspond to the Minkowski embedding, whereas the four continuous curves on the bottom correspond to the black hole embedding. The four dotted curves in the middle correspond to the critical embedding, in which the branes touch the horizon at a single point.}
        \label{plong}
    \end{center}
\end{figure}

The inversion needed to express $u$ as a function of $\rho$ must be performed numerically for generic values of the anisotropy, due to the fact that the function ${\cal F}$ entering in (\ref{defro2}) is generically not known analytically. Moreover, the equations to be solved are highly non-linear, which also requires a numerical approach. This can be readily done, resulting in the plots of Fig. \ref{plong}, where the different profiles for the embeddings are represented. Analytic solutions are however possible in the limits of high and low temperature and are detailed in App.~\ref{appB}. 

Notice that the anisotropy has the effect of contributing to bending the brane towards the horizon:  increasing the anisotropy of the background, the bending of the branes toward the horizon also increases. Moreover, anisotropic profiles converge to the flat geometry at larger distances. The critical embedding corresponds to a greater value of the asymptotic distance to the brane, holographically related to $M_\mt{q}/T$. This means that the critical temperature $T_\mt{c}$ is smaller in the anisotropic case. We can then already see that {\it increasing the value of the anisotropy must have a similar effect as increasing the temperature of the black hole}. We shall comment further upon this point after the computation of the free energy.


\section{Thermodynamics of the flavor D7-branes}
\label{sec4}

In order to study the thermodynamics of the flavor D7-branes, we need to evaluate their on-shell action for both embeddings. This will give the corresponding free energy. As usual, the on-shell action suffers from divergences due to the infinite volume of AdS, which need to be regularized using holographic renormalization \cite{reviewHR,Karch:2005ms}, as we do next.


\subsection{Holographic renormalization}
\label{secHR}

In what follows we shall work in Einstein frame, which is obtained by rescaling the metric (\ref{metric}) by a factor $e^{-\frac{1}{2}\phi}$:
\begin{equation}
 ds^{2}=\frac{e^{-\frac{1}{2}\phi}}{u^{2}}\left(-\mathcal{FB}dt^{2}+dx^{2}+dy^{2}+e^{-\phi}dz^{2}+\frac{du^{2}}{\mathcal{F}}\right)
 +d\Omega_{5}^{2}\,.
\label{metrica_10dim}
\end{equation}
This has the advantage of making the metric a direct product of an asymptotically AdS space with a round sphere. We also transform the radial coordinate $u$ to the Fefferman-Graham (FG) coordinate $v$ given by \cite{MT2}
\be
v=u-\frac{1}{12}a^2u^3+\left(-\frac{b_4+7 f_4}{56}+\frac{1091}{32256} a^4-\frac{1}{16}a^4\log u\right)u^5+{\cal O}(u^7)\,,
\ee
in terms of which the fields have the following near-boundary expansions
\begin{eqnarray}
\phi&=&-\frac{1}{4}a^{2}v^{2}+\left(\frac{2}{7}b_{4}-\frac{47}{4032}a^{4}-\frac{1}{6}a^{4}\log{v}\right)v^{4}+\mathcal{O}(v^{6})\,,
\cr
\mathcal{F}&=&1+\frac{11}{24}a^{2}v^{2}+\left(f_{4}+\frac{11}{144}a^{4}+\frac{7}{12}a^{4}\log{v}\right)v^{4}+\mathcal{O}(v^{6})\,,
\cr
\mathcal{B}&=&1-\frac{11}{24}a^{2}v^{2}+\left(b_{4}-\frac{11}{144}a^{4}-\frac{7}{12}a^{4}\log{v}\right)v^{4}+\mathcal{O}(v^{6})\,,
\label{fields_v}
\end{eqnarray}
and the metric (\ref{metrica_10dim}) becomes 
\begin{equation}
 ds^{2}=\frac{dv^2}{v^2}+\frac{1}{v^2}g_{ij}dx^{i}dx^{j}+d\Omega_5^2,
\label{fondo}
\end{equation} 
with
\begin{eqnarray}
 g_{tt}&=&-1+\frac{1}{24}a^{2}v^{2}+\left(\frac{2749}{16128}a^{4}-\frac{23}{28}b_{4}-\frac{3}{4}f_{4}+\frac{1}{24}a^{4}\log{v}\right)v^{4}+\mathcal{O}(v^{6}),
\cr
 g_{xx}&=&g_{yy}=1-\frac{1}{24}a^{2}v^{2}+\left(\frac{71}{1792}a^{4}-\frac{5}{28}b_{4}-\frac{1}{4}f_{4}-\frac{1}{24}a^{4}\log{v}\right)v^{4}+\mathcal{O}(v^{6}),
\cr
 g_{zz}&=&1+\frac{5}{24}a^{2}v^{2}+\left(\frac{1163}{16128}a^{4}-\frac{13}{28}b_{4}-\frac{1}{4}f_{4}+\frac{1}{8}a^{4}\log{v}\right)v^{4}+\mathcal{O}(v^{6}).
\label{g_zz}
\end{eqnarray}
In these expressions $f_4$ and $b_4$ are two coefficients that are not determined from the asymptotic equations of motion, as expected, but that can be extracted once a particular solution is known (either analytically in limiting regimes of small or large temperature or numerically for intermediate regimes \cite{MT2}).

A further ingredient needed to continue the analysis is the boundary expansion of the embedding profile $\psi=\cos\vartheta$ introduced above. This can be obtained from the action
\begin{equation}
 S_\mt{D7}=-T_\mt{D7}\textrm{vol}(S^3)\int d^4x\, dv\,
 \frac{e^{\phi}}{v^4}\sqrt{-\text{det}(g_{ij})(1-\psi^{2})^{3}\left(\frac{1}{v^2}+\frac{\psi'^{2}}{1-\psi^{2}}\right)}
\label{SD7}
\end{equation}
and from the equation of motion resulting from it. In this formula, the prime denotes a derivative with respect to $v$. The asymptotic solution for $\psi$ turns out to be given by \cite{Jahnke:2013rca}\footnote{The coefficients $m$ and $c$ in this formula are related to the ones in \cite{Jahnke:2013rca} as follows: $m=\psi_1$ and $c=\psi_3+\frac{a^2}{12}\psi_1$. }
\begin{equation}
\psi=mv+\left(c+\frac{5}{24}a^{2}m\log{v}\right)v^{3}+\mathcal{O}(v^{5}).
\label{psi_v}
\end{equation}
The parameters $m$ and $c$ are related to the quark masses and to the chiral condensate \cite{thermobrane}. Using the expansions (\ref{g_zz}), integrating in $v$, and evaluating the action $S_\mt{D7}$ at a cutoff $v=\epsilon\to 0$, one finds that it contains the following divergences:
\begin{eqnarray}
\frac{S_\text{div}}{T_\mt{D7}\textrm{vol}(S^3)}&= & -\int d^{4}x\left(\frac{1}{4\epsilon^{4}}-\left(\frac{5}{48}a^{2}+\frac{1}{2}m^{2}\right)\frac{1}{\epsilon^{2}}
+\frac{1}{12}a^{4}\log^{2}{\epsilon}
\right.
\cr && \hskip 2cm \left.
+\left(\frac{47}{4032}a^{4}-\frac{5}{12}a^{2}m^{2}-\frac{2}{7}b_{4}\right)\log{\epsilon}
+\mathcal{O}(\epsilon^{0})\right).
\label{particular_divergences}
\end{eqnarray}
The task is now to cancel these divergences with appropriate counterterms and to extract the finite value of the action.  In App.~\ref{series}, it is shown that (\ref{particular_divergences}) can be rewritten to get a more covariant appearance as
\bea
\frac{S_\text{div}}{T_\mt{D7}\textrm{vol}(S^3)}&=&  -\int d^{4}xe^{\phi_{(0)}}\sqrt{-g_{(0)}}\left(\frac{a_{(0)}}{\epsilon^{4}}+\frac{a_{(1)}}{\epsilon^{2}}
+(a_{(2)}-\phi_{(4)})\log{\epsilon}+a_{(3)}\log^{2}{\epsilon}+\mathcal{O}(\epsilon^{0})\right)\,,\cr&&
\label{divergences}
\eea
with
\bea
&& a_{(0)}=\frac{1}{4},
\cr
&&
a_{(1)}=-\frac{1}{4\epsilon^{2}}\left(\frac{5e^{2\phi}}{12}\partial^{i}\chi\partial_{i}\chi+2\psi^{2}\left(1-\frac{5e^{2\phi}}{12}\partial^{i}\chi\partial_{i}\chi\log{\epsilon}\right)\right),
\cr
&&
 a_{(2)}=-\frac{5e^{2\phi}}{12\epsilon^4}\psi^2\partial^{i}\chi\partial_{i}\chi\left(1-\frac{e^{2\phi}}{16}\partial^{i}\chi\partial_{i}\chi\log{\epsilon}\right),
\cr
&&
a_{(3)}=\frac{e^{4\phi}}{12\epsilon^{4}}(\partial^{i}\chi\partial_{i}\chi)^{2},
\label{as}
\eea
and 
\bea
e^{\phi_{(0)}}\sqrt{-g_{(0)}}&=&\epsilon^{4}\sqrt{-\gamma}e^{\phi} \cr && \times \left(1+\frac{5e^{2\phi}}{24}\partial^{i}\chi\partial_{i}\chi+\frac{e^{4\phi}}{6}(\partial^{i}\chi\partial_{i}\chi)^{2}\left(\log{\epsilon}-\frac{1}{16}\right)-\frac{e^{6\phi}}{144}(\partial^{i}\chi\partial_{i}\chi)^{3}\log{\epsilon}\right)\,,\cr &&
\label{phi_det}
\eea
where $g_{(0)}=\det (g_{(0)ij})$ and $\gamma=\det(\gamma_{ij})$ is the determinant of the metric on the surface $v=\epsilon$. 

Plugging \eqref{as}-\eqref{phi_det} into \eqref{divergences} and discarding finite terms suggests to consider the following counterterms
\begin{equation}
\begin{split}
 \frac{S^{(a)}_{\text{ct}}}{T_\mt{D7}\textrm{vol}(S^{3})}=&\frac{1}{4}\int d^{4}x\sqrt{-\gamma}e^{\phi}\left(1-\frac{5e^{2\phi}}{24}\partial^{i}\chi\partial_{i}\chi-2\psi^{2}\right)\\
 &+\frac{1}{24}\log{\epsilon}\int d^{4}x\sqrt{-\gamma}e^{3\phi}\left(e^{2\phi}(\partial^{i}\chi\partial_{i}\chi)^{2}-5\psi^{2}\partial^{i}\chi\partial_{i}\chi\right)\\
 &+\frac{1}{12}\log^{2}{\epsilon}\int d^{4}x\sqrt{-\gamma}e^{5\phi}(\partial^{i}\chi\partial_{i}\chi)^{2}.
\end{split}
\label{contraterminos}
\end{equation}

We still need to take care of the term proportional to $\phi_{(4)}$. It is, of course, not possible to express $\phi_{(4)}$ solely in terms of $\phi$, so that one has to consider certain combinations of fields to remove the associated logarithmic divergence. Noticing that $\phi_{(4)}$ appears in
\begin{equation}
\int d^{4}x\sqrt{-\gamma}e^{\phi}=\int d^{4}x\left(\frac{1}{\epsilon^{4}}-\frac{5}{24\epsilon^{2}}a^{2}-\frac{1}{6}a^{4}\log{\epsilon}+\phi_{(4)}+\mathcal{O}(\epsilon^{2})\right),
\label{b}
\end{equation}
one sees that a possible counterterm is given by
\begin{equation}
 \frac{S_\text{ct}^{(b)}}{T_\mt{D7}\textrm{vol}(S^3)}=-\log{\epsilon}\int d^{4}x\sqrt{-\gamma}e^{\phi}\,.
\end{equation}
This term introduces extra divergences, which can, however, be eliminated by a third counterterm
\bea
 \frac{S^{(c)}_{\text{ct}}}{T_\mt{D7}\textrm{vol}(S^3)}&=& \log{\epsilon}\int d^{4}x \sqrt{-\gamma}\left(1-\frac{e^{2\phi}}{4}\partial^{i}\chi\partial_{i}\chi-\frac{7e^{5\phi}}{48}(\partial^{i}\chi\partial_{i}\chi)^{2}\right)\cr
 && -\frac{1}{6}\log^{2}{\epsilon}\int d^{4}x\sqrt{-\gamma}e^{5\phi}(\partial^{i}\chi\partial_{i}\chi)^{2}.
\eea

The full counterterm action is then given by the sum of all the pieces above
\be
S_\text{ct}=S_\text{ct}^{(a)}+S_\text{ct}^{(b)}+S_\text{ct}^{(c)}\,,
\label{Sctfinal}
\ee
and reads
\bea
\frac{S_\text{ct}}{T_\mt{D7}\textrm{vol}(S^3)}
&=&
\frac{1}{4}\int d^{4}x\sqrt{-\gamma}e^{\phi}\left(1-\frac{5e^{2\phi}}{24}\partial^{i}\chi\partial_{i}\chi-2\psi^{2}\right)\cr
&& +\log{\epsilon}\int d^{4}x \sqrt{-\gamma}\left(1-e^{\phi}-\frac{e^{2\phi}}{4}\partial^{i}\chi\partial_{i}\chi\left(1+\frac{5e^{\phi}}{6}\psi^{2}\right)-\frac{5e^{5\phi}}{48}(\partial^{i}\chi\partial_{i}\chi)^{2}\right)
\cr&&
 -\frac{1}{12}\log^{2}{\epsilon}\int d^{4}x\sqrt{-\gamma}e^{5\phi}(\partial^{i}\chi\partial_{i}\chi)^{2}.
\eea
Notice that for $\phi=\chi=0$, this reduces, as it should, to
\begin{equation}
 \frac{S_\text{ct}}{T_\mt{D7}\textrm{vol}(S^{3})}= \frac{1}{4}\int d^{4}x\sqrt{-\gamma}(1-2\psi^{2})\,,
\end{equation}
which is, up to finite terms, the expression found in \cite{thermobrane} for the isotropic case. 

With the addition of these counterterms to the divergent action, we obtain a regularized action $S_\mt{D7}^\text{reg}=S_\text{div}+S_\text{ct}$, which is finite and reads
\begin{equation}
 \begin{split}
\frac{S_\mt{D7}^{\text{reg}}}{T_\mt{D7}\textrm{vol}(S^{3})}=& \int d^{4}x\left(\frac{1}{14}b_{4}-c\,m+\frac{5}{48}a^{2}m^{2}+\frac{241}{5376}a^{4}+\mathcal{O}(\epsilon)\right)\,.
\end{split}
\end{equation}

A few comments are in order at this point. First of all, we note that {\it a priori} there could be other scheme-dependent finite counterterms that would have to be included in (\ref{Sctfinal}), see \cite{Karch:2005ms}. These counterterms are in fact vanishing in our case, since neither the metric nor the embedding profiles depend on the boundary coordinates. The axion does depend on $z$, but its field strength does not.

Second, there could in principle be a contribution from the conformal anomaly of the background \cite{MT2}, which enters at order ${\cal O}(a^4)$, but this does not affect the phase transition between the Minkowski and black hole embeddings, since it only contributes to an additive overall shift to the free energy.

Finally, notice that the chiral condensate can be obtained from varying the renormalized action with respect to the quark mass and reads
\be
\vev{\bar q q}=c-\frac{5}{24} a^2 m
\label{qc}
\,.
\ee
This expression is exact in the anisotropy parameter $a$. The chiral condensate was also computed in \cite{thermobrane}, where the authors showed that its value vanishes when $T/M_q$ approaches zero, as expected in an anomaly free gauge theory, like the one dual to the gravitational set up used in that work. The behavior of the condensate can be modified if the theory has an anomaly. To see that this can be the case, let us remember the Gell Mann-Oakes-Renner relation \cite{GellMann:1968rz}, that in a construction like ours, where all the quark masses are equal to $M_q$, can be written as
\be
f_\pi {m_\pi}^2=2M_q \vev{\bar q q}\,, 
\label{gmor}
\ee
with $m_\pi$ being the mass of the pion and $f_\pi$ the pion decay rate, which is different from zero if the chiral currents are not conserved, that is, in the presence of an anomaly.

It is clear that the nature of the $m_\pi$ plays a relevant role here, and even if the meson spectrum has not been computed for our anisotropic background, in the context of the gauge/gravity correspondence with flavor \cite{Kruczenski:2003be}, the mass of the pion, just like $M_q$, is expected to be proportional to the parameter $m$. From these considerations and the non vanishing $f_\pi$, given in our case by the anomaly in the gauge theory, the relation (\ref{gmor}) suggests that $ \vev{\bar q q}\sim m$, which is consistent with (\ref{qc}).
We have numerically verified that for $a=0$ (resulting in $f_\pi=0$) we recover a vanishing $\vev{\bar q q}$ as $M_q/T$ goes to infinity. For non vanishing $a$, we have confirmed the monotonic growth of the condensate as the mass of the quark gets much larger than the temperature. This check was done numerically for values of $M_q/T$ with non vanishing $a$ which were not too large, since the integrand in (\ref{SD7}) grows as we get farther from the horizon, compromising numerical stability. A detailed study of the behavior of the condensate in this limit would require a deeper understanding of the vacuum and the meson spectrum in this anisotropic theory, which is an interesting problem beyond the scope of this current investigation.


\subsection{Free energy}

At this point, we have all the ingredients to compute the free energy of the branes and determine which are the energetically favorable configurations. The free energy is notoriously given by the on-shell Euclidean action, which we now compute.

Switching back to the $u$ coordinate and expanding all the fields around the boundary to compute the counterterm action, one can show after some trivial albeit tedious manipulations that the regularized action becomes\footnote{Here we have included a term that goes like $\psi^4$, which was considered in \cite{thermobrane}.}
\begin{equation}
\begin{split}
\frac{S_\mt{D7}^{\text{reg}}}{T_\mt{D7}\textrm{vol}(S^{3})}=&\frac{1}{T}\left(G(m)-\frac{355}{32256}a^{4}+c\, m-\frac{a^{2}m^{2}}{48}-\frac{5}{56}b_{4}-\frac{1}{8}f_{4}-\frac{1}{4}m^{4} -\frac{1}{4\uh^{4}}\right. \\
& \left. +\frac{a^2+24m^2}{48\uh^{2}}-\log{\uh}\left(\frac{299a^{4}}{4032}-\frac{5a^{2}m^{2}}{12}-\frac{2b_{4}}{7}\right)-\frac{1}{12}a^{4}\log^{2}{\uh}\right),
\end{split}
\label{FreeEnergy}
\end{equation}
where the integral
\begin{equation}
\begin{split}
G(m)=\int_{u_{\text{boundary}}}^{\uh}du&\left(\frac{e^{-\frac{3}{4}\phi}}{u^{5}}(1-\psi^{2})\sqrt{\mathcal{B}(1-\psi^{2}+e^{\frac{1}{2}\phi}u^{2}\mathcal{F}\psi'^{2})}+\frac{a^{4}}{6u}\log{u}
\right. \\& \left.+\frac{1}{u^3}\left(\frac{a^2}{24}+m^{2}\right)  +\frac{1}{u}\left(\frac{299a^{4}}{4032}-\frac{5a^{2}m^{2}}{12}-\frac{2b_{4}}{7}\right)-\frac{1}{u^5}\right)
\end{split}
\label{Gm}
\end{equation}
is finite by construction. The free energy is related to this action by
\begin{equation}
F=T  \, S_\mt{D7}^{\text{reg}}.
\end{equation}

Notice that in the expressions (\ref{FreeEnergy})-(\ref{Gm}) above there appear various geometrical parameters that originate from the expansions of the fields and are unphysical quantities. We would like instead to find a final result only in terms of the physical observables $a$, $T$, and the quark mass $M_\mt{q}$. To this scope, we need to find the explicit map between the two sets of parameters
\be
(\uh, \phi_\mt{H}, {\cal B}_\mt{H}, m, \ldots) \quad \to \quad (a, T, M_\mt{q})\,.
\ee
First of all, we notice that since the flavor branes are probes in a fixed geometry, the relation between $T$ and $(\uh,\phi_\mt{H},{\cal B}_\mt{H})$ has to be the same as in the case without the branes \cite{MT2}, namely (\ref{temperature}). The same is true for $a$, which is also fixed once we fix the background \cite{MT2}. What is then left to do is to extract the dependence of $M_\mt{q}$ on these parameters.

Finally, we can plot \eqref{FreeEnergy} as a function of the various parameters. Out of the dimensionful parameters $a$, $T$, and $M_\mt{q}$ we can form two dimensionless ratios, which we pick to be $a/M_\mt{q}$ and $T/M_\mt{q}$.\footnote{In practice, in all the numerical evaluations needed to obtain our plots, we have fixed the position of the horizon to be $\uh=1$.  As mentioned in the Introduction, this results in freezing the conformal anomaly scale $\mu$ to a fixed value.} The former choice has the advantage of displaying the similarity between anisotropy and temperature, while the latter choice allows for a more immediate comparison with the isotropic results present in the literature. Note that the background itself is not invariant under simultaneous rescaling of $a$ and $T$ even if the ratio $a/T$ is kept fixed \cite{MT2}. Consequently we do not expect our results to be invariant under a similar rescaling, even if the adimensional variables are kept fixed. Nonetheless, we explore a range of values for the parameters $T$, $a$ and $M_\mt{q}$ and even if the particulars change, the general behavior is the same, and the same conclusions can be reached.

\begin{figure}[ht!]
 \centering
\includegraphics[width=.7\textwidth]{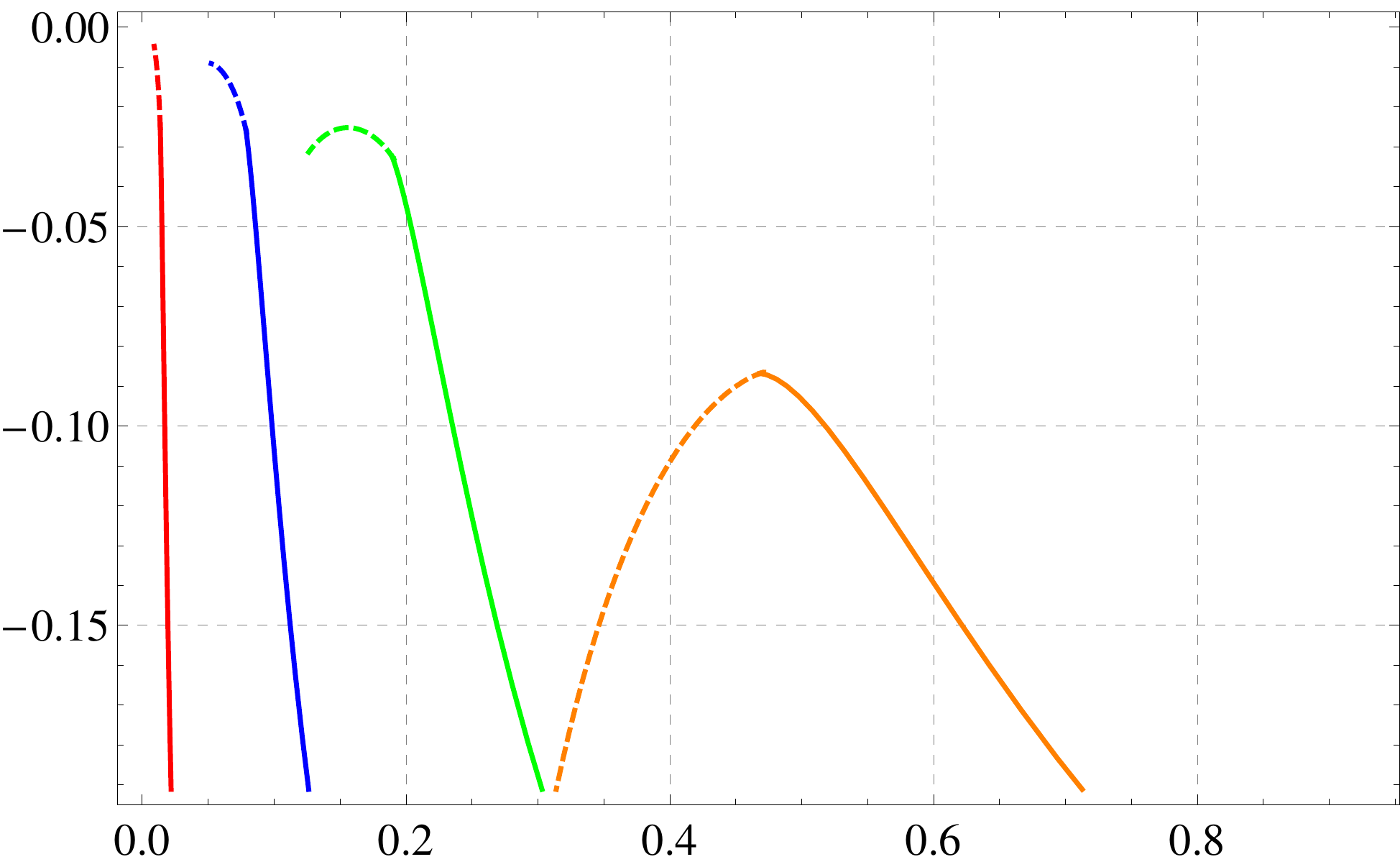}
  \put(-18,-10){$\frac{a}{M\mt{q}}$}
   \put(-320,160){ $\frac{F}{T K}$}
\caption{Free energy as a function of $a/M_\mt{q}$ for fixed values of $T/a$. From left to right, we have fixed $T/a=$25.3 (red), 4.38 (blue), 1.81 (green), and 0.72 (orange). The dashed part of the lines represents the Minkowski embedding, whereas the continuous part represents the black hole embedding. Here $K=T_\mt{D7}\mbox{vol}(S^3)\mbox{vol}(x)\nf$.}
\label{free-en1}
\end{figure}

\begin{figure}
\begin{center}
\begin{tabular}{cc}
\setlength{\unitlength}{1cm}
\hspace{-0.9cm}
\includegraphics[width=7cm]{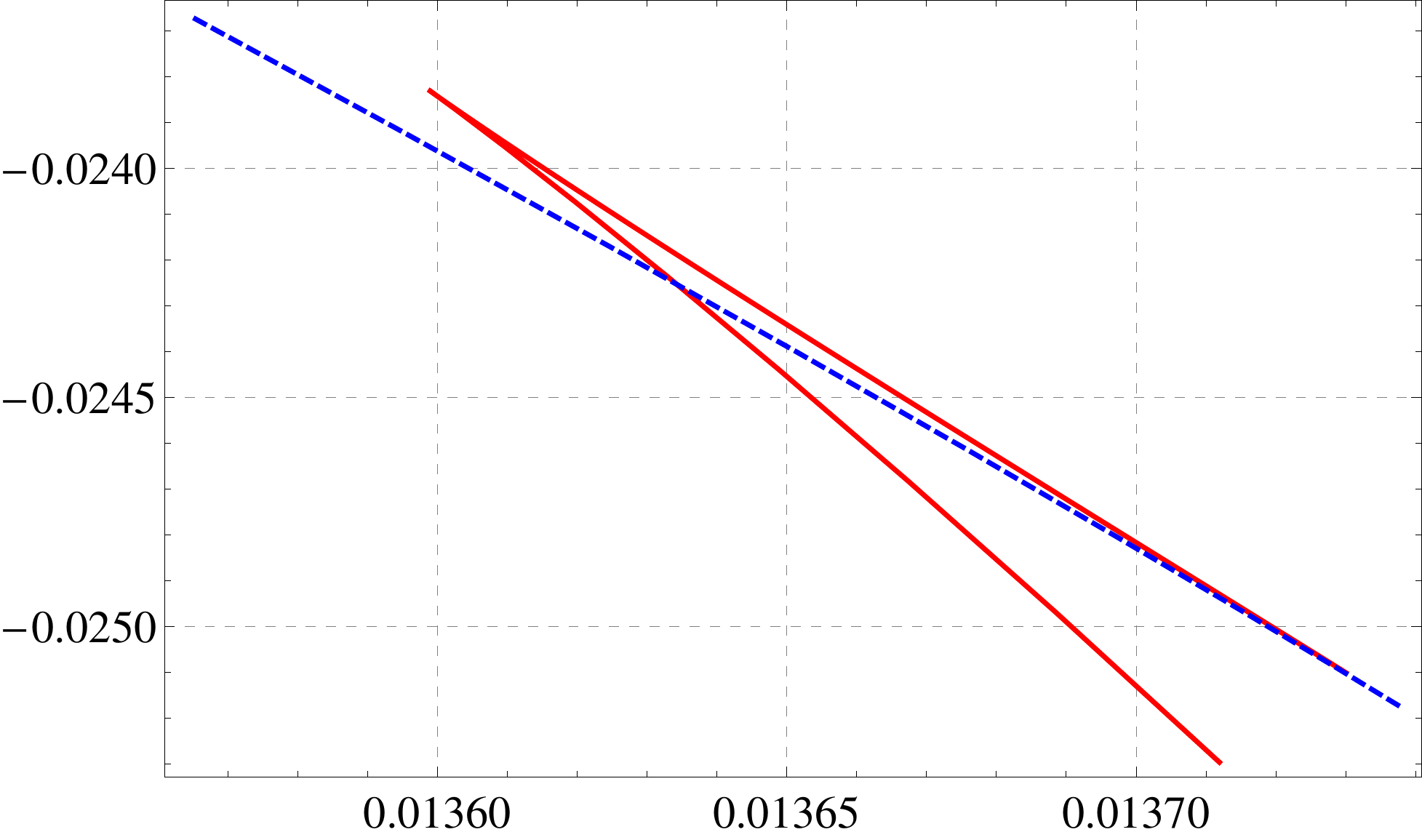} 
\qquad\qquad & 
\includegraphics[width=7cm]{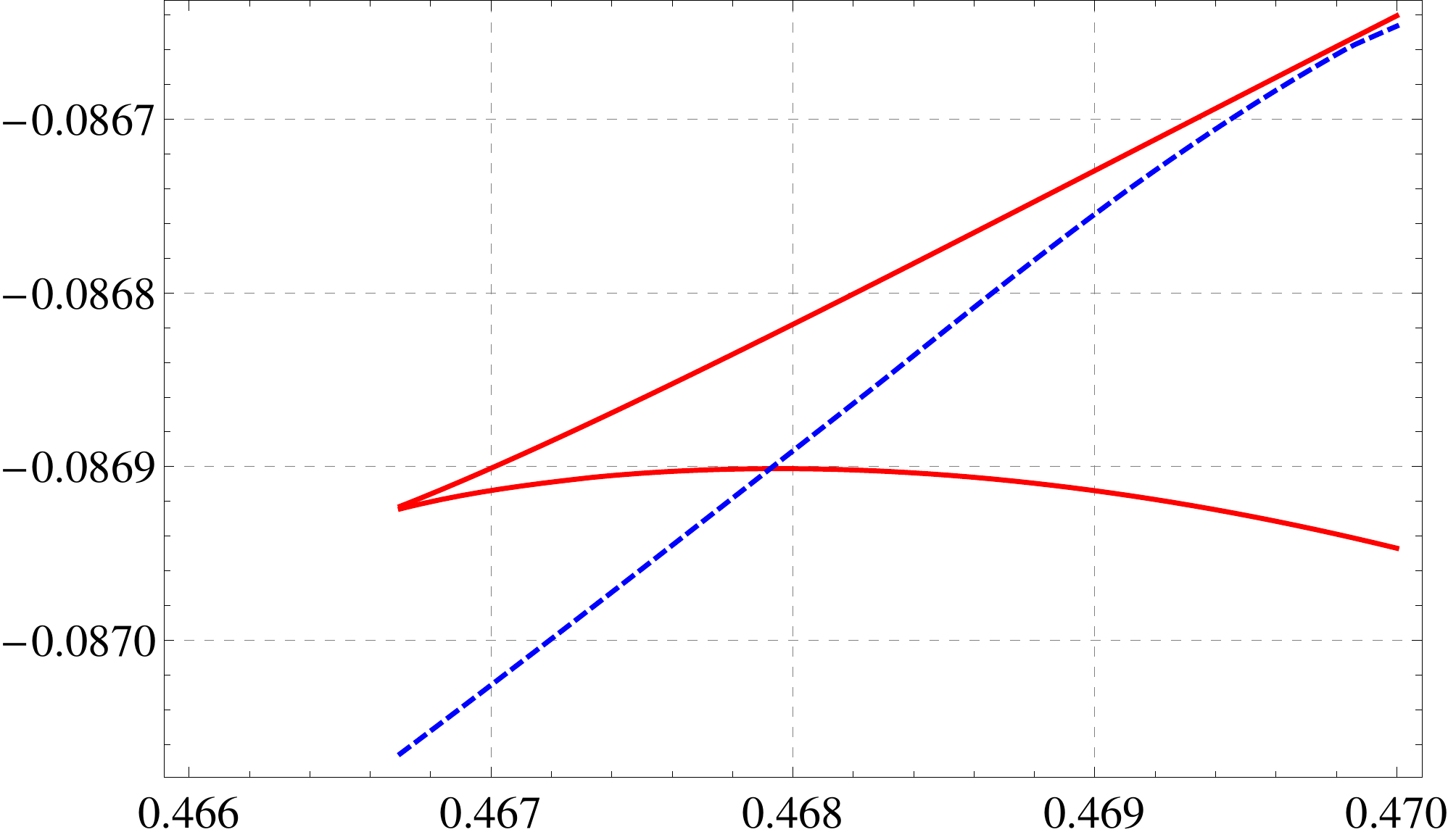}
\qquad
 \put(-450,70){\rotatebox{90}{$F/TK$}}
   \put(-250,-10){$\frac{a}{M_\mt{q}}$}
    \put(-218,70){\rotatebox{90}{$F/TK$}}
   \put(-18,-10){$\frac{a}{M_\mt{q}}$}
 \\
(a) & (b)\\
& \\
\hspace{-0.9cm}
\includegraphics[width=7cm]{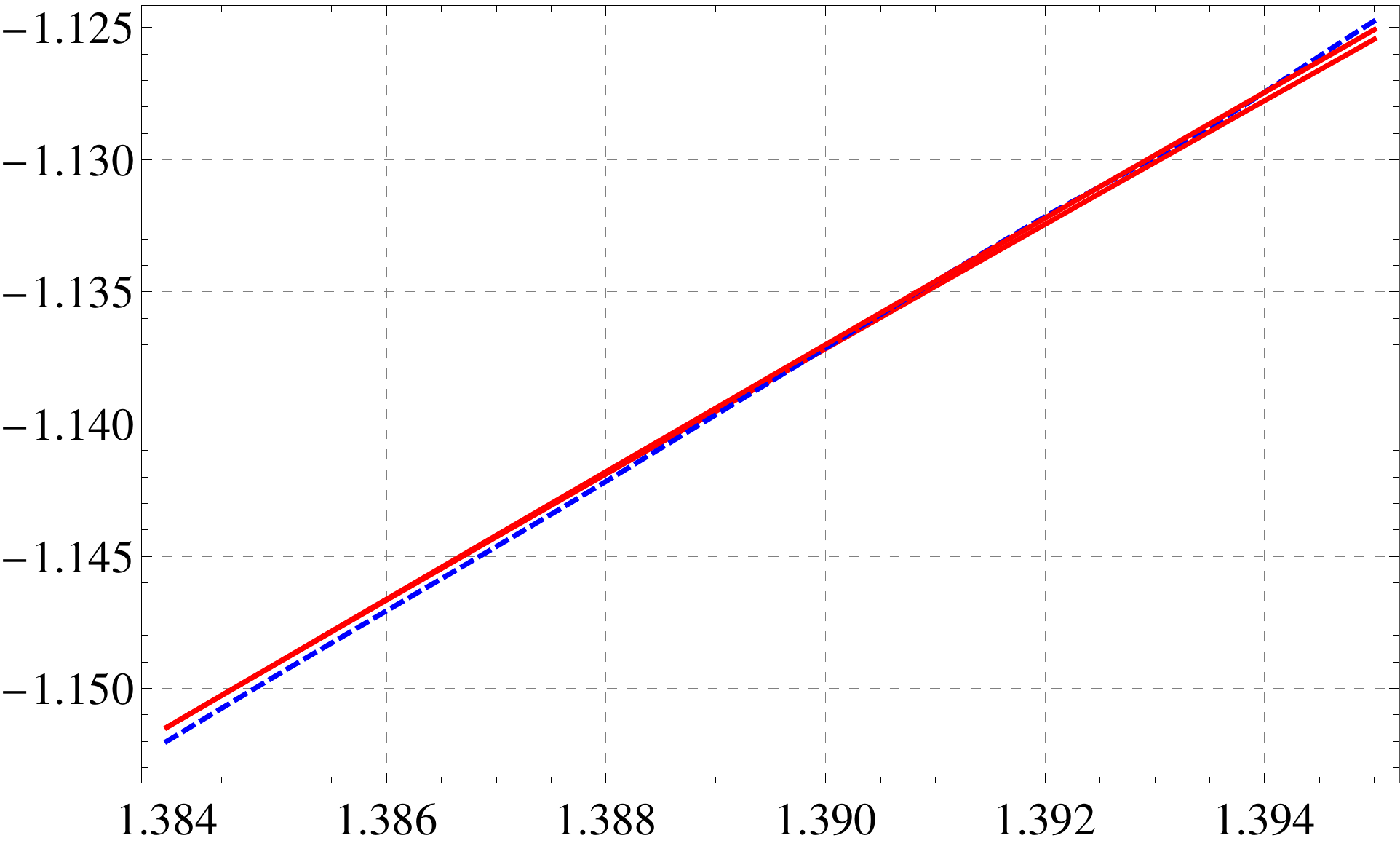} 
\qquad\qquad & 
\includegraphics[width=7cm]{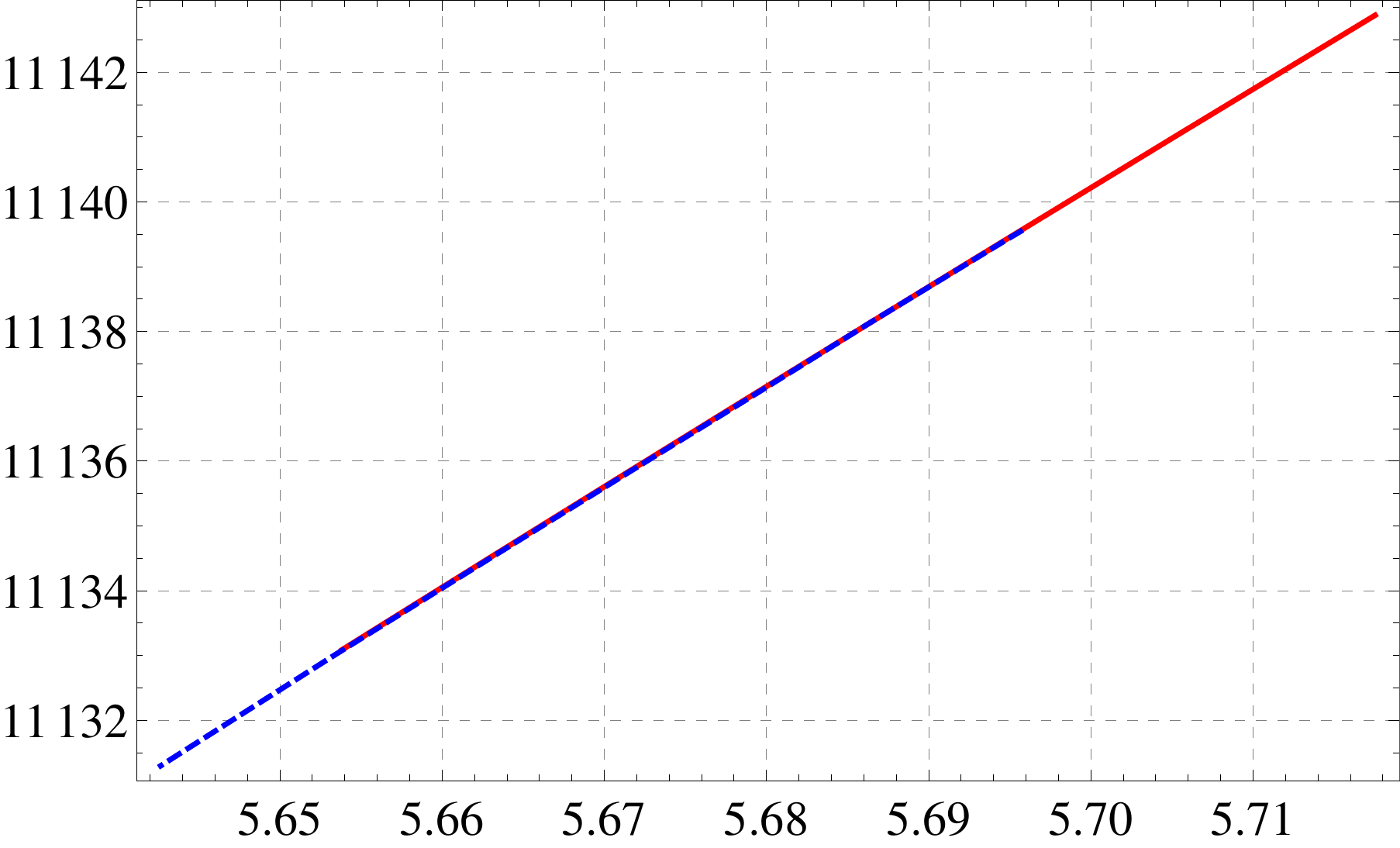}
\qquad
 \put(-450,70){\rotatebox{90}{$F/TK$}}
   \put(-250,-10){$\frac{a}{M_\mt{q}}$}
    \put(-218,70){\rotatebox{90}{$F/TK$}}
   \put(-18,-10){$\frac{a}{M_\mt{q}}$}
         \\
(c)& (d) 
\end{tabular}
\end{center}
\caption{\small Zoom into the transition region between the Minkowski and black hole embeddings. The values of the temperature are $T/a=$ 25.3 (a), 0.72 (b), 0.22 (c), and 0.04 (d). The blue, dashed line represents the Minkowski embedding, whereas the solid, red line represents the black hole embedding. Here $K=T_\mt{D7}\mbox{vol}(S^3)\mbox{vol}(x)\nf$.}
\label{zooms}
\end{figure}

\begin{figure}[ht!]
 \centering
\includegraphics[width=.6\textwidth]{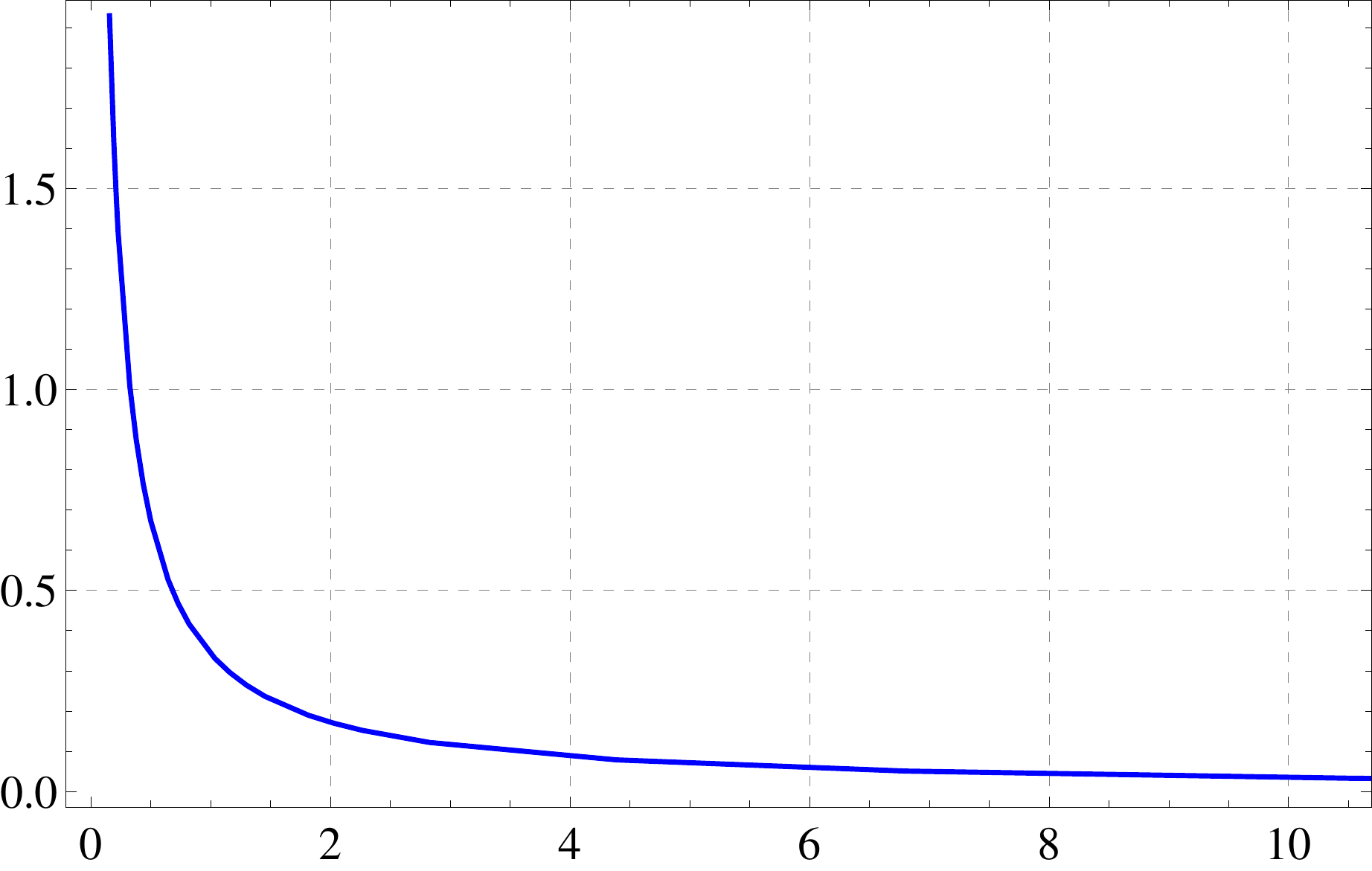}
 \put(-13,-10){$\frac{T}{a}$}
   \put(-280,140){ $\frac{a_\mt{c}}{M_\mt{q}}$}
\caption{Critical anisotropy as a function of the temperature.}
\label{critanis}
\end{figure}

In Fig.~\ref{free-en1}, we plot the free energy as a function of $a/M_\mt{q}$ for different values of $T/a$. The free energy is divided by $T \,K$, where  $K=T_\mt{D7}\mbox{vol}(S^3)\mbox{vol}(x)\nf$ scales like $T^3$, making the quantity on the vertical axis adimensional. In order to better understand the order of the phase transition, we zoom in Fig.~\ref{zooms} over the parameter region in which the transition takes place. At small anisotropy and large temperature, the plots display the ``swallow tail'' structure (Fig.~\ref{zooms} (a) and (b)), which is typical of first order phase transitions, with a discontinuous first order derivative of the curve. This is in agreement with the findings of \cite{thermobrane}. Increasing the anisotropy over temperature, however, the swallow tail disappears, smoothing out the curve (Fig.~\ref{zooms} (c) and (d)). We discover then an interesting consequence of the introduction of this extra parameter in the system: the order of the transition between the Minkowski and black hole phases depends on the value of the anisotropy and grows with it.
From Fig.~\ref{free-en1}, it is also obvious that the critical value of $a/M_\mt{q}$ at which the transition takes place increases as we decrease $T/a$. This critical anisotropy as a function of the temperature is detailed in Fig.~\ref{critanis}.

\begin{figure}[ht!]
 \centering
 \includegraphics[width=.7\textwidth]{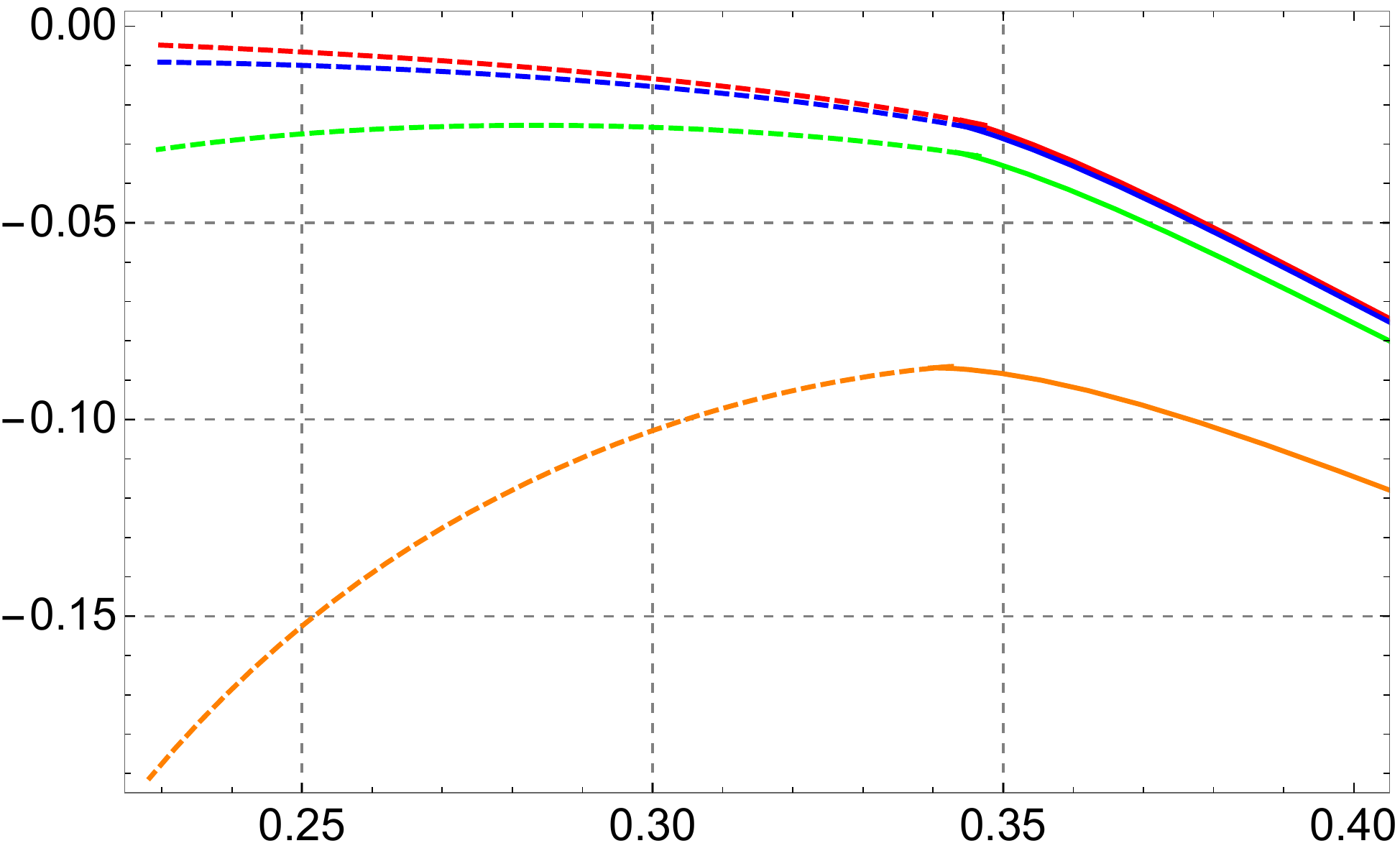}
   \put(-18,-10){$\frac{T}{M_\mt{q}}$}
   \put(-320,160){ $\frac{F}{T K}$}
\caption{Free energy as a function of $T/M\mt{q}$ for fixed values of $a/T$. From top to bottom, we have fixed $a/T=$0.03 (red), 0.22 (blue), 0.55 (green), and 1.37 (orange). The dashed part of the lines represents the Minkowski embedding, whereas the continuous part represents the black hole embedding. Here $K=T_\mt{D7}\mbox{vol}(S^3)\mbox{vol}(x)\nf$. Notice how increasing the value of the anisotropy from (a) to (d), the discontinuity in the derivative of the curves smooth out, signalling an increase of the order of the phase transition.}
\label{free-en2}
\end{figure}

\begin{figure}
\begin{center}
\begin{tabular}{cc}
\setlength{\unitlength}{1cm}
\hspace{-0.9cm}
\includegraphics[width=7cm]{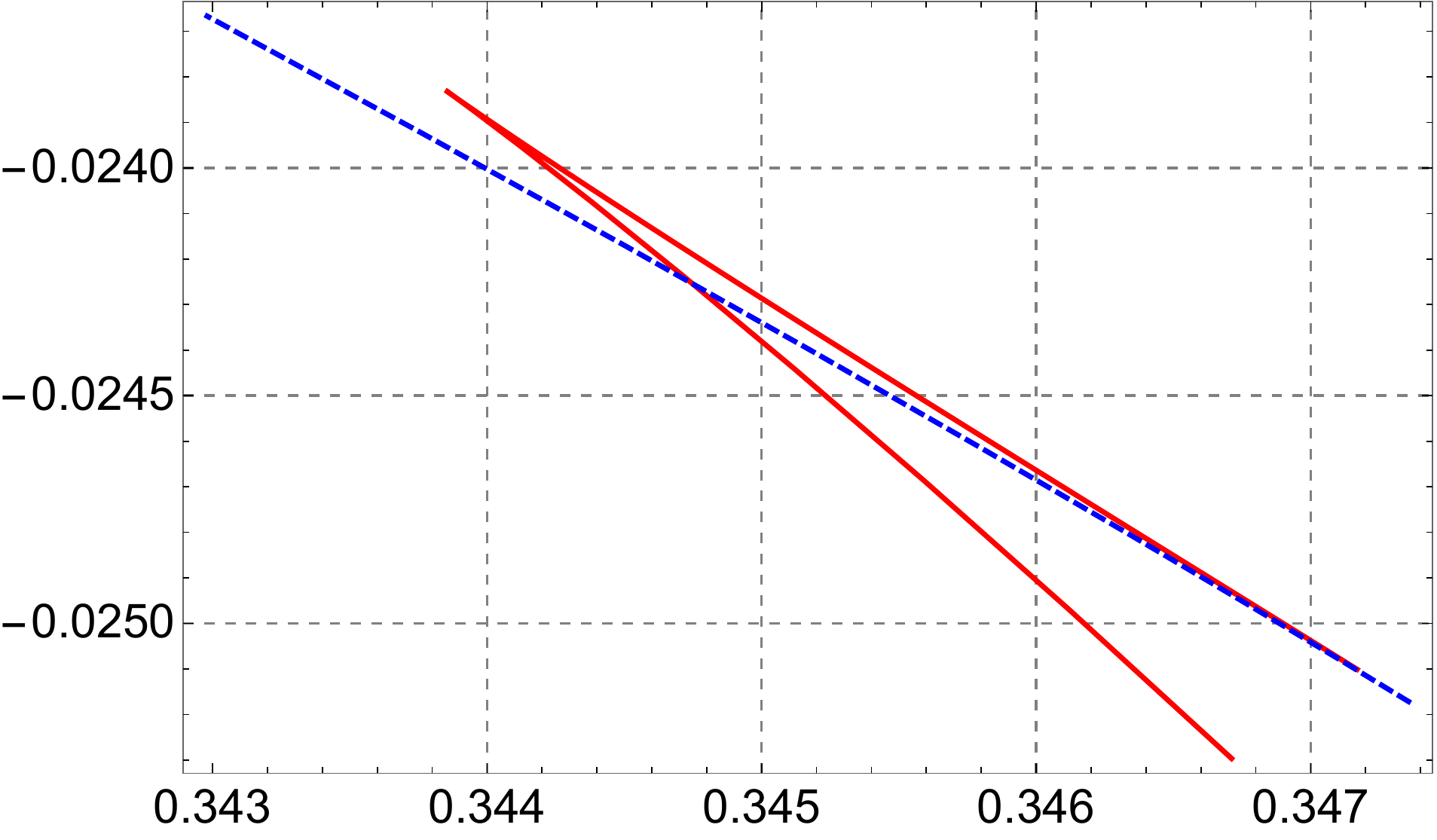} 
\qquad\qquad & 
\includegraphics[width=7cm]{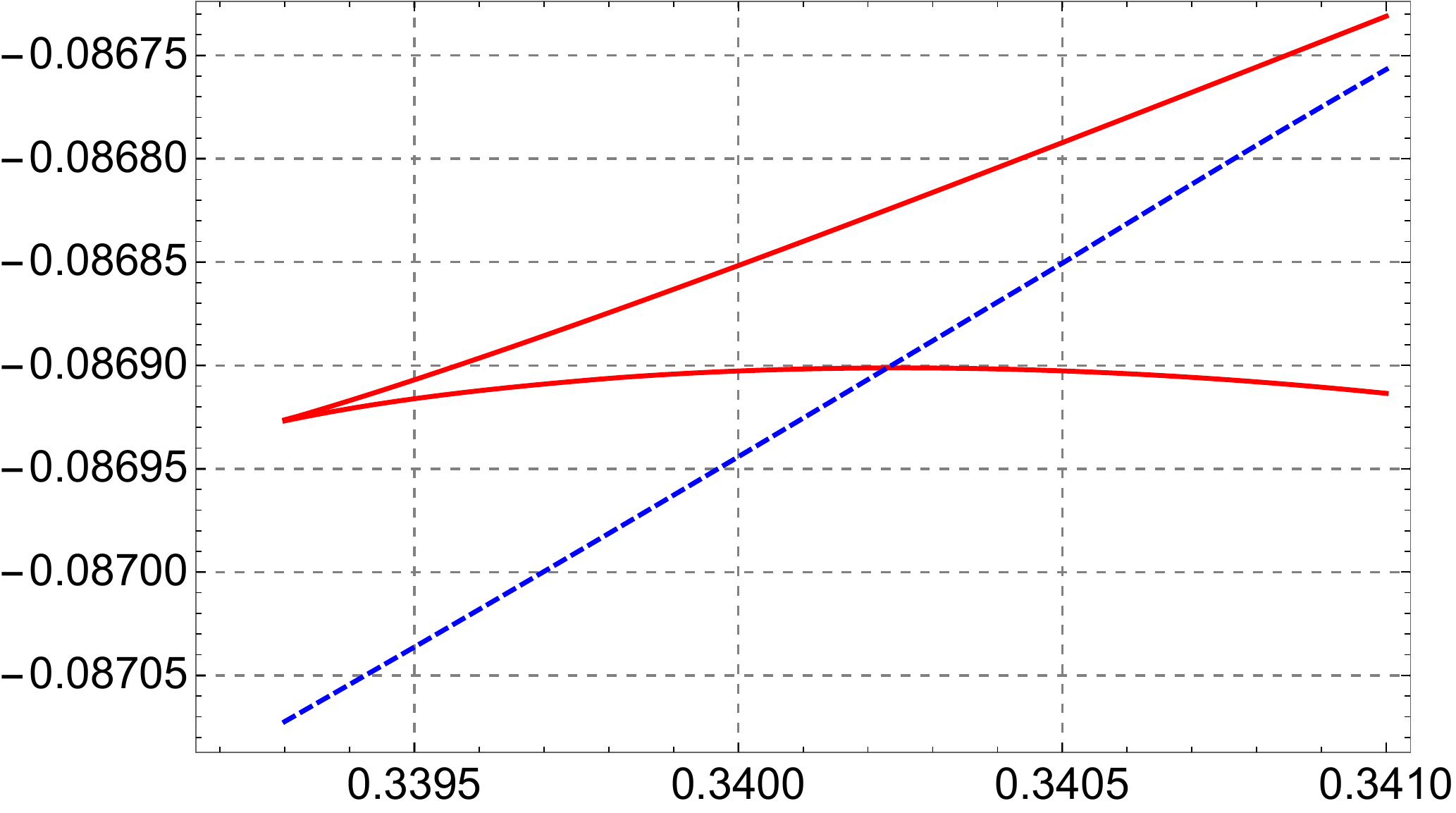}
\qquad
 \put(-450,70){\rotatebox{90}{$F/TK$}}
   \put(-250,-10){$\frac{T}{M_\mt{q}}$}
    \put(-218,70){\rotatebox{90}{$F/TK$}}
   \put(-18,-10){$\frac{T}{M_\mt{q}}$}
 \\
(a) & (b)\\
& \\
\hspace{-0.9cm}
\includegraphics[width=7cm]{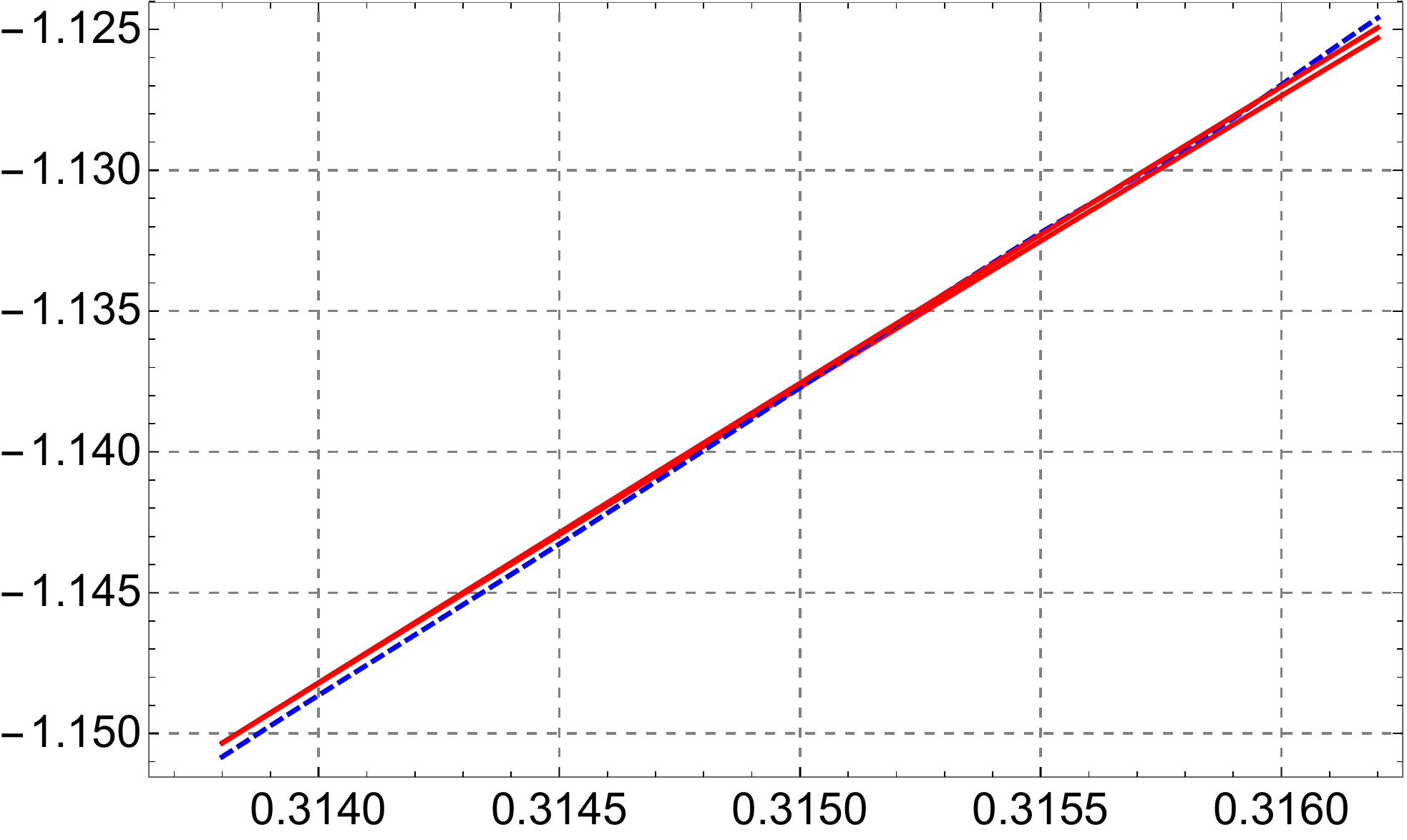} 
\qquad\qquad & 
\includegraphics[width=7cm]{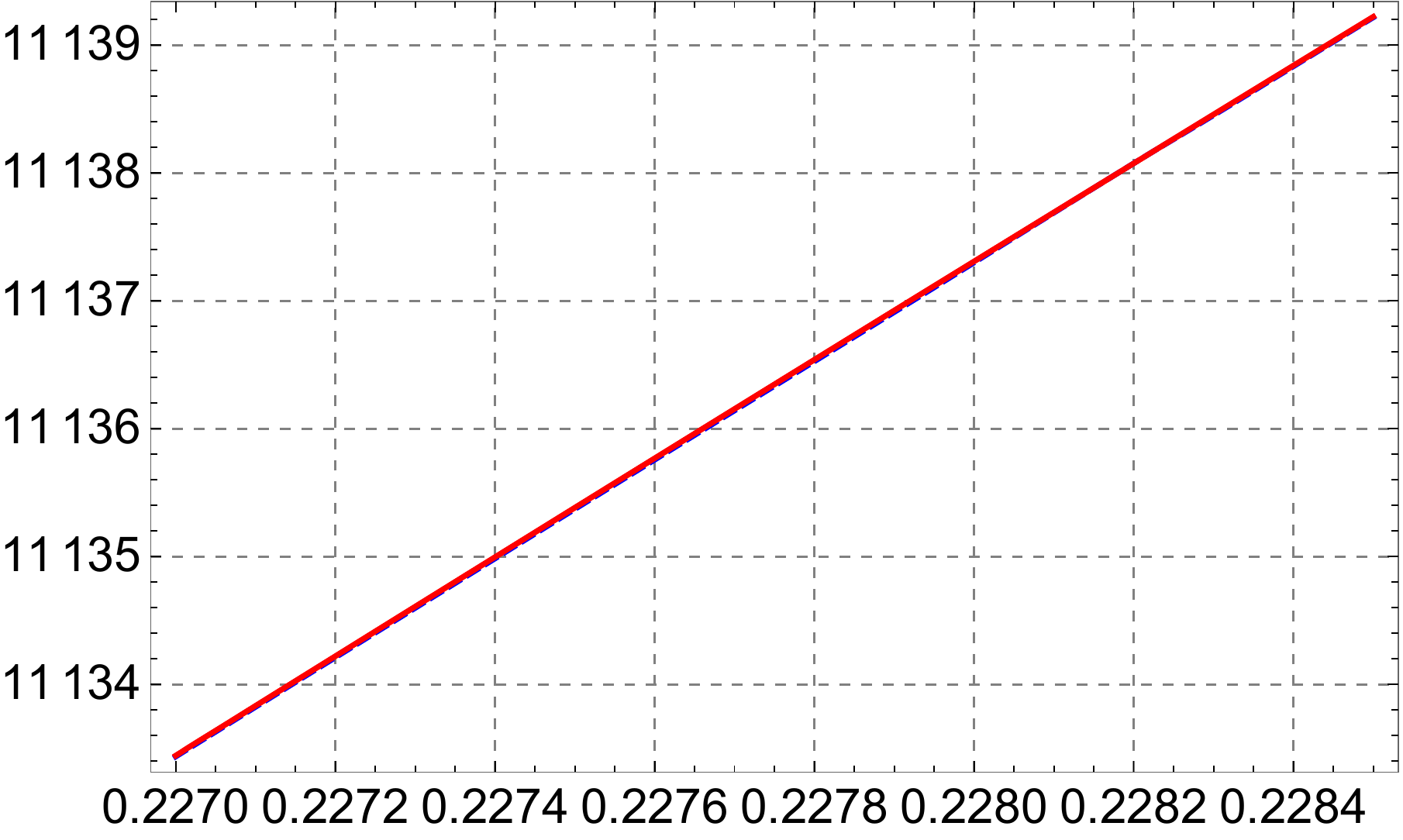}
\qquad
 \put(-450,70){\rotatebox{90}{$F/TK$}}
   \put(-250,-10){$\frac{T}{M_\mt{q}}$}
    \put(-218,70){\rotatebox{90}{$F/TK$}}
   \put(-18,-10){$\frac{T}{M_\mt{q}}$}
         \\
(c)& (d) 
\end{tabular}
\end{center}
\caption{\small Zoom into the transition region between the Minkowski and black hole embeddings. The values of the anisotropy are $a/T=$ 0.03 (a), 1.37 (b), 4.41 (c), and 24.9 (d). The blue, dashed line represents the Minkowski embedding, whereas the solid, red line represents the black hole embedding. Here $K=T_\mt{D7}\mbox{vol}(S^3)\mbox{vol}(x)\nf$. As in Fig.~\ref{zooms}, we see that increasing the anisotropy from (a) to (d) smooths out the transition region of the curves.}
\label{zooms2}
\end{figure}

\begin{figure}[ht!]
 \centering
\includegraphics[width=.6\textwidth]{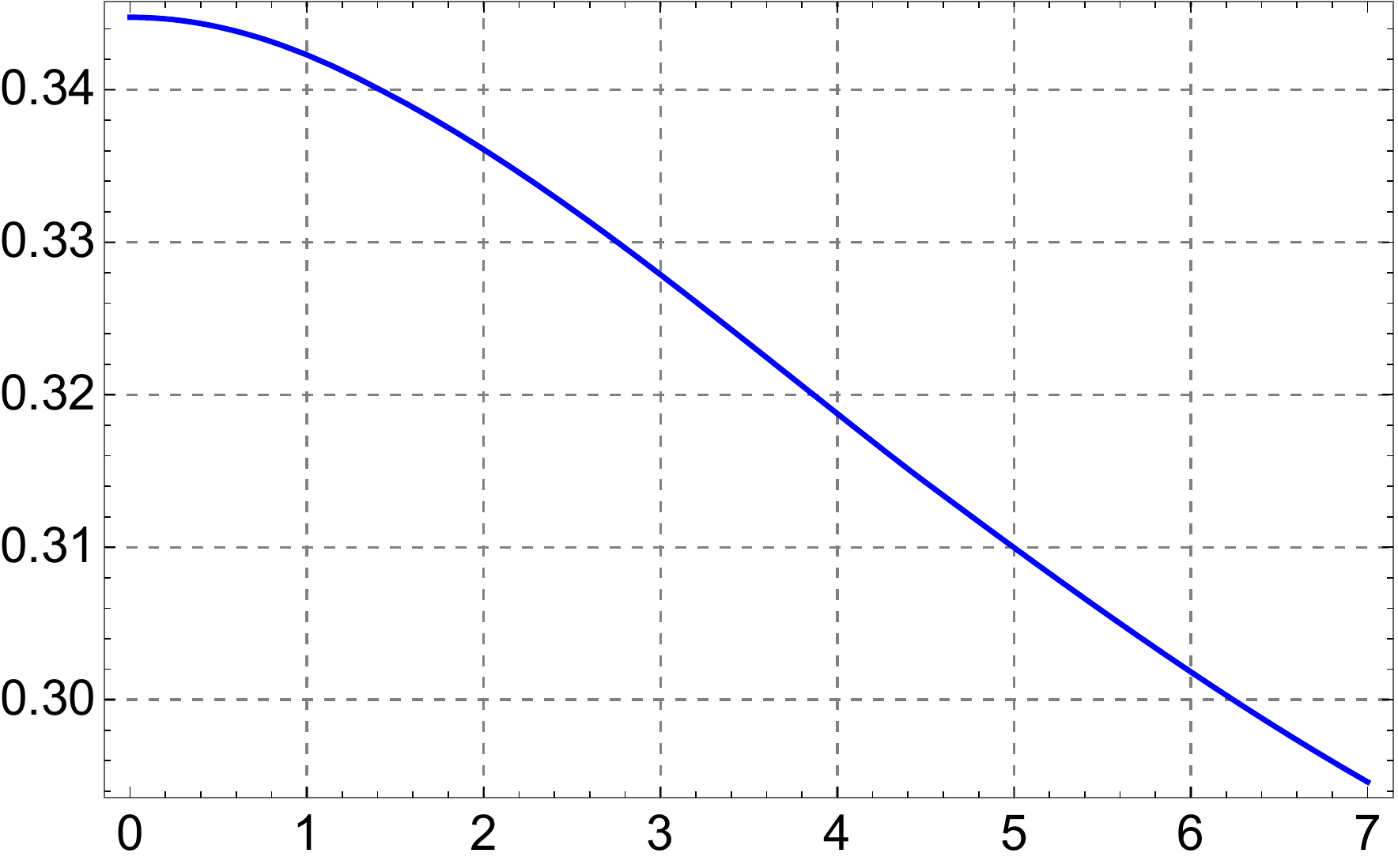}
 \put(-13,-10){$\frac{a}{T}$}
   \put(-280,140){ $\frac{T\mt{c}}{M_\mt{q}}$}
\caption{Critical temperature as a function of the anisotropy.}
\label{critT}
\end{figure}

In Fig.~\ref{free-en2}, we plot the free energy as a function of $T/M\mt{q}$ for fixed values of $a/T$. In Fig.~\ref{zooms2} we zoom over the region of the phase transition and we see again a very similar behavior as the one in Fig.~\ref{zooms}: increasing the anisotropy has the effect of smoothing out the transition which, starting from first order for small $a/T$ (Fig.~\ref{zooms2} (a) and (b)), becomes of higher order (Fig.~\ref{zooms2} (c) and (d)). 
The critical temperature as a function of the anisotropy is reported in Fig.~\ref{critT}. Comparing with Fig.~\ref{critanis} we see that the decrease of the critical temperature with the increase of the anisotropy is much slower than the decrease of the critical anisotropy with the increase of the temperature in that plot.


\subsection{Entropy}

\begin{figure}[ht!]
 \centering
 \includegraphics[width=.7\textwidth]{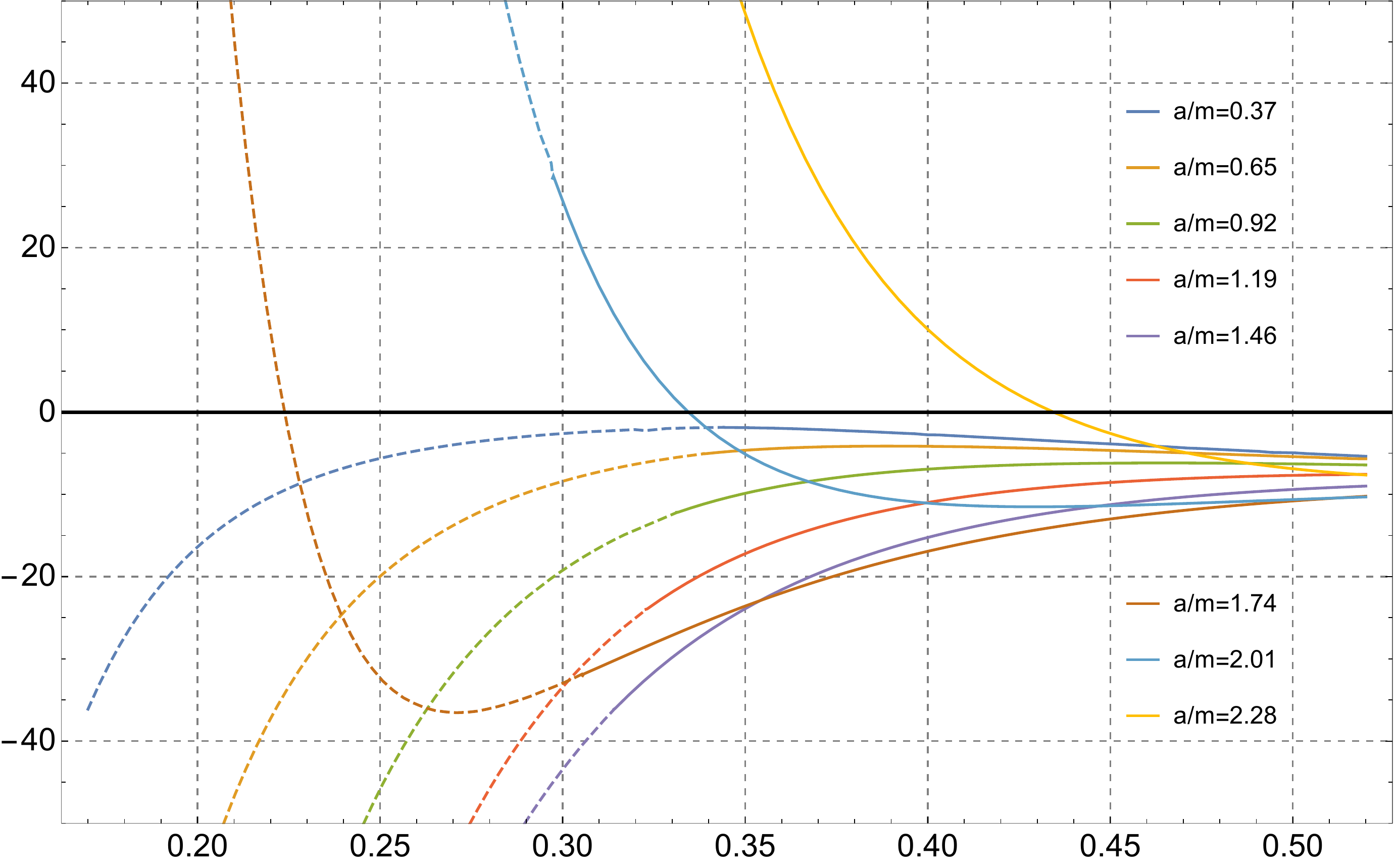}
  \put(-20,68){$\frac{T}{M_\mt{q}}$}
  \put(-320,140){\rotatebox{90}{ $ F/T^4$}}
\caption{Free energy as a function of the temperature for $a/M_\mt{q}$ fixed. The decreasing curves (in yellow, blue and brown) have, from left to right, $a/M_\mt{q}=1.74, 2.01, 2.28$. The increasing curves have, from left to right, $a/M_\mt{q}=0.37, 0.65, 0.92, 1.19, 1.46$. As usual, the dashed regions of the curves correspond to the Minkowski embedding, whereas the solid regions correspond to the black hole embeddings.}
\label{ent-1}
\end{figure}

\begin{figure}[ht!]
 \centering
\includegraphics[width=.7\textwidth]{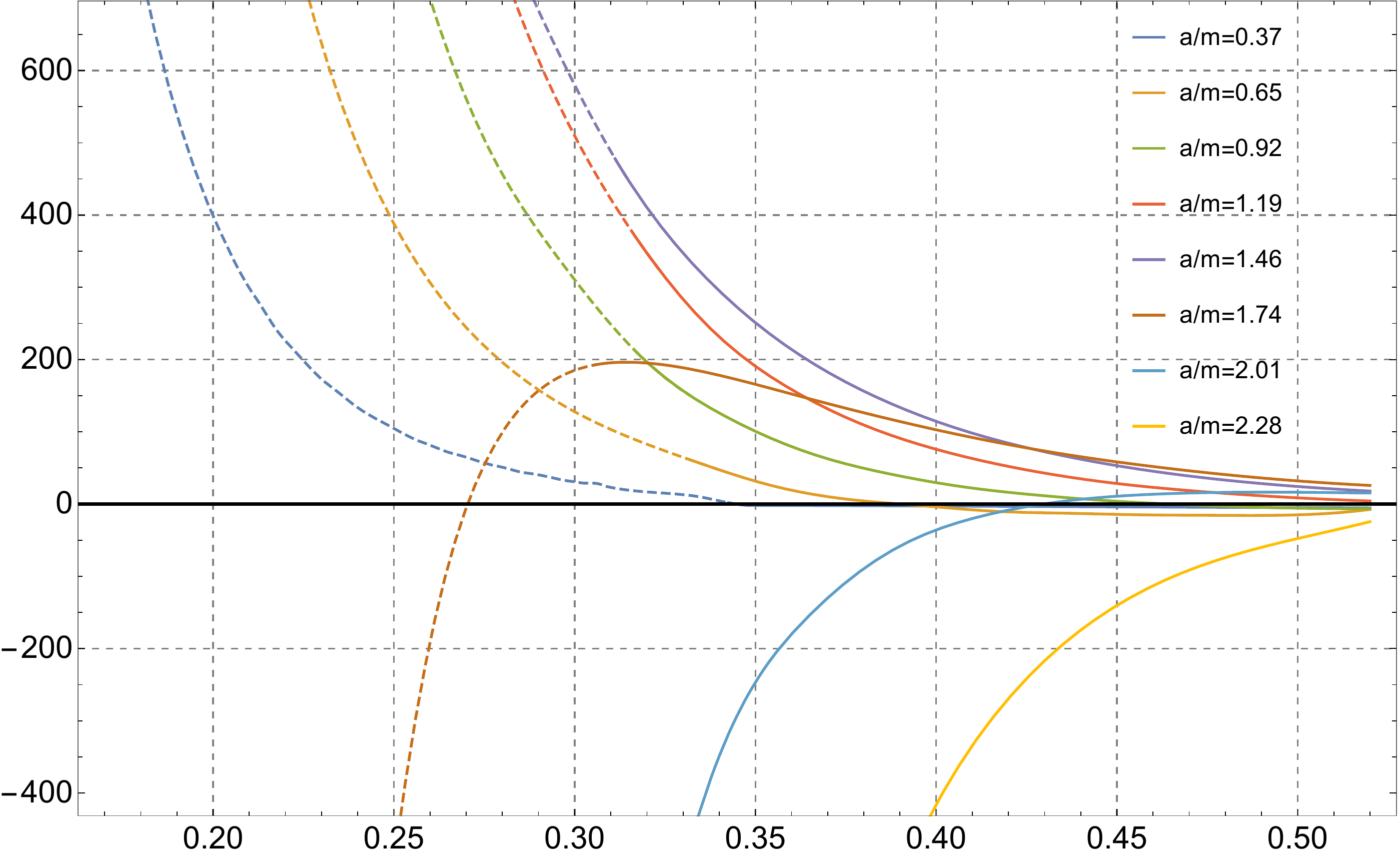}
 \put(-20,52){$\frac{T}{M_\mt{q}}$}
  \put(-320,140){\rotatebox{90}{ $ s/T^4$}}
\caption{Entropy density as a function of the temperature for $a/M_\mt{q}$ fixed.  The increasing curves (in yellow, blue and brown) have, from left to right, $a/M_\mt{q}=1.74, 2.01, 2.28$. The decreasing curves have, from left to right, $a/M_\mt{q}=0.37, 0.65, 0.92, 1.19, 1.46$. As usual, the dashed regions of the curves correspond to the Minkowski embedding, whereas the solid regions correspond to the black hole embeddings.}
\label{ent-2}
\end{figure}

Another interesting thermodynamical quantity is the entropy density, which is given by taking the derivative of the free energy with respect to the temperature
\begin{equation}
S=-\frac{\partial F}{\partial T}\,,
\end{equation}
while maintaining $a/M_\mt{q}$ fixed. In order to take this derivative, we need to recompute the free energy while maintaining this new ratio fixed. The results for the free energy and the entropy density are plotted in Figs.~\ref{ent-1}-\ref{ent-2}.


\section{Discussion}
\label{concl}

In this paper we have studied fundamental matter phase transitions in a strongly coupled plasma with a spatial anisotropy, using holography. This has been done by considering the embedding of flavor D7-branes in a type IIB supergravity solution with a non-vanishing, anisotropy-inducing axion field. In a heavy ion collision experiment, the anisotropic direction would correspond to the beam direction. The flavor branes allow for the introduction of degrees of freedom in the fundamental representation of the gauge group, which we have called quarks with a slight abuse of language. We have considered the so-called quenched flavor approximation, in which the number of quark flavors $\nf$ is much smaller than the number of colors $\nc$, in order to have probe branes in a fixed background geometry. It is important to notice that these quarks are allowed to have finite mass (which is controlled by the radial position of the D7-branes) and are therefore dynamical particles. In this regard, this work extends the analysis done in \cite{melting}, where the physics of infinitely heavy mesons was investigated. 

We have focused on the transition which happens between a phase in which the D7-branes lie outside of the black hole horizon present in the background and a phase in which they fall into the horizon. The corresponding physics in the dual gauge theory side is that of a transition between a discrete meson spectrum with a mass gap and a continuous, gapless spectrum \cite{thermobrane}. It is important to stress that the transition studied here is not a confinement/deconfinement transition (as the background geometry always contains a black hole horizon), but a transition of fundamental matter, which can survive in a bound state even if the surrounding plasma is deconfined. 

Interestingly, the order of the transition depends on the degree of anisotropy of the system, with higher anisotropies making the transition smoother. It would be interesting to determine precisely the order of the phase transition and the critical exponents as functions of the parameter $a$. Complex critical exponents signal a first-order transition, while real exponents a higher order transition. To compute these quantities, a near-horizon scaling analysis along the lines of the one performed in \cite{thermobrane,yaffe} woule be required. 

From a technical point of view, a novelty of this paper is represented by the renormalization procedure that we have adopted in order to compute the free energy. There appear divergent terms that go like a logarithm squared of the cutoff, which have been eliminated with  an appropriate set of counterterms. We have also obtained an analytic exact expression for the chiral condensate as a function of the anisotropy parameter. However, the main general lesson we can draw from our computations is that, besides a critical temperature, there is also a critical anisotropy at which the transition can happen, at fixed temperature. In fact, the anisotropy seems to play a very similar role as the temperature: increasing the anisotropy bends the D7-branes more and more toward the black hole horizon. This confirms the findings of \cite{melting}, where the anisotropy was seen as being a possible mechanism for the dissociation of heavy quarkonium.

This can be heuristically understood as the brane being `pulled' more strongly into the horizon due to the anisotropy. This has to do with the gravitational strength at the particular position in which the brane is located. In the isotropic case, there is only one way in which the gravitational pull changes along the bulk, so increasing the gravitational pull at a certain position implies increasing it in all the bulk. The non-trivial axion gives cause for new behavior, because it has the effect of making the gravitational pull change more drastically, in such a way that the branes get to feel the same gravitational pull `earlier', that is, while being located further from the horizon. The reason why $a/m$ acts similarly to $T/m$ is that in the end, this has the same effect in the phase transition as the gravitational pull being increased overall, which is what happens when the temperature is increased. On the field theory side, this effect is harder to understand. Previously \cite{melting}, we showed that the anisotropy produces an effective force in the preferred direction. One way to understand why mesons would be easier to melt is that the existence of different forces in different directions has the combined effect of pulling them apart more easily.
 
One possible extension of this analysis, that goes beyond the scope of this paper, would be the computation of the meson spectrum. Mesons, i.e. pairs of quarks and antiquarks, are described by the endpoints of strings ending on the D7-branes and their spectrum is given by the fluctuations around these branes in the Minkowski embedding. This analysis was carried over in the isotropic case in \cite{thermobrane}, confirming the general intuition that in the Minkowski phase the spectrum is gapped and discrete. Relatedly, the screening length for infinitely heavy mesons was already studied in \cite{melting} and could be repeated in this setup for mesons with finite masses. In the present case, repeating these computations would be complicated by the fact that the background geometry is not generically known analytically. It would however be interesting in order to compare the stability and screening length for mesons propagating in different directions of the plasma with the real-world results obtained in experiments such as \cite{Abelev:2013ila}.


\section*{Acknowledgements}

We are happy to thank David Mateos for collaboration at the early stages of this work complemented by later discussions, and Timo Alho and Matt Luzum for discussions. DA and LP acknowledge partial support from DGAPA-UNAM Grant No. IN113115. DF is supported by an Alexander von Humboldt Foundation fellowship. DT acknowledges partial financial support from CNPq and from the FAPESP grants 2014/18634-9 and 2015/17885-0. He also thanks ICTP-SAIFR for their support through FAPESP grant 2016/01343-7.


\appendix
 
\section{Near-horizon expansions}
\label{AppA} 

\begin{figure}[h!]
    \begin{center}
        \includegraphics[width=0.45\textwidth]{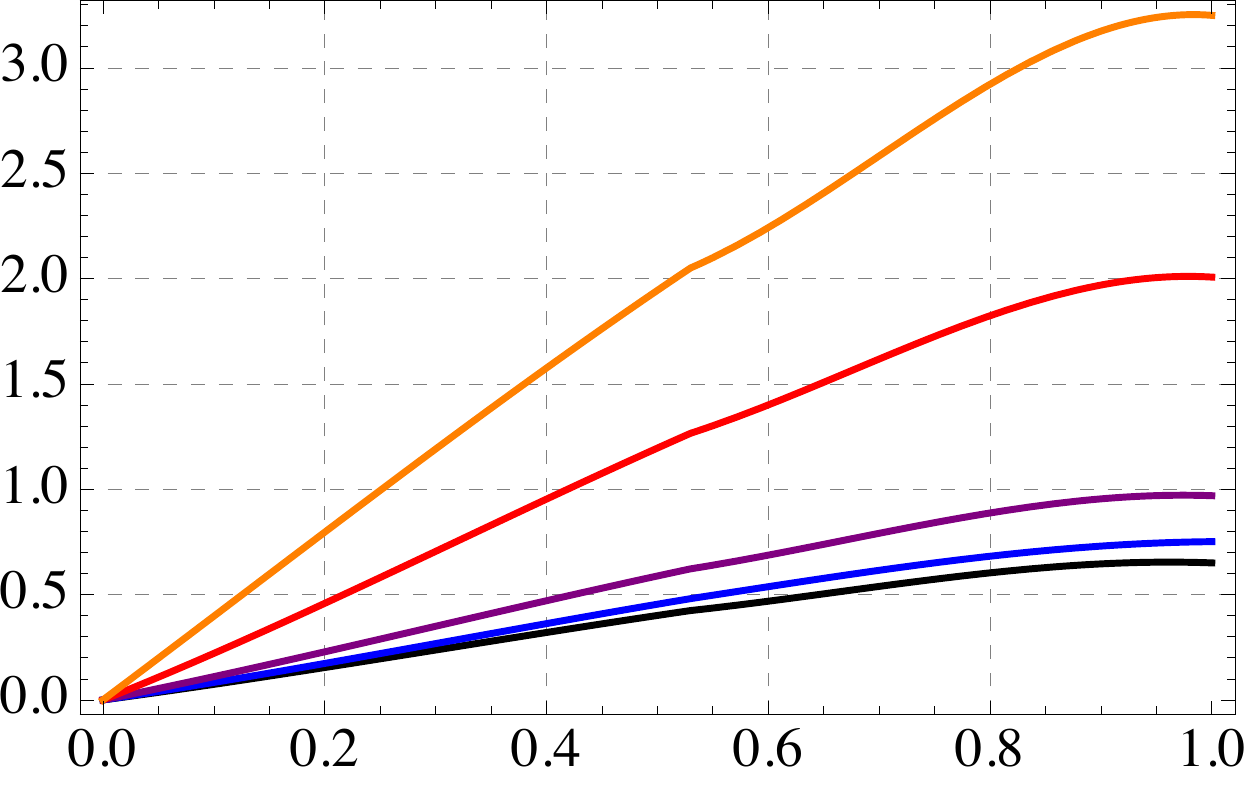}
         \put(-250,100){$M_\mt{q}/\sqrt{\lambda} T$}
         \put(10,5){$\psi_\mt{H}$}
        \caption{The curves correspond, from bottom to top, to $a/T=0,\, 4.41,\, 12.2,\, 86,\, 249$. }
        \label{plotpsi}
    \end{center}
\end{figure}

In order to obtain numerically the embedding function $\psi$, it is necessary to impose boundary conditions at the horizon and integrate outward from there. To this purpose, the following near-horizon expansions for the dilaton and metric fields are needed \cite{MT2}
\bea
\phi(\rho) &=& \phi_\mt{H}+ \frac{e^{\frac{7 \phi_\mt{H}}{2}} }{4} {u_\mt{H}}^2 (\rho-1)^2 + {\cal O} \left( {(\rho-1)}^3 \right)\,,\cr
\cf(\rho) &=& \frac{1}{64} e^{-\frac{\phi_\mt{H}}{2}} {\left( 16 + e^{\frac{7 \phi_0}{2}} {u_\mt{H}}^2 \right)}^2 (\rho-1)^2
+ {\cal O} \left( {(\rho-1)}^3 \right)\,,\cr
\cb(\rho) &=& 1 +
\left( 14 - \frac{5}{8} e^{\frac{7 \phi_\mt{H}}{2}} {u_\mt{H}}^2 - \frac{224}{16 + e^{\frac{7 \phi_\mt{H}}{2}} {u_\mt{H}}^2} \right) (\rho-1)^2
+ {\cal O} \left( {(\rho-1)}^3 \right)\,.
\eea
The function $P(\rho)$ that enters into the equation for the embedding gets expanded as
\be
P(\rho) = \frac{e^{-\frac{\phi_\mt{H}}{2}}}{8 {u_\mt{H}}^4} \sqrt{e^{-\frac{\phi_\mt{H}}{2}} {\left( 16 + e^{\frac{7 \phi_\mt{H}}{2}} {u_\mt{H}}^2 \right)}^2}
\left( 1 - \frac{9}{2} (\rho-1) \right) (\rho-1)
+ {\cal O} \left( {(\rho-1)}^3 \right)\,,
\label{pro}
\ee
resulting in the following expression for the embedding function
\bea
\psi(\rho) &=& \psi_\mt{H} - \frac{3}{4} \psi_\mt{H} (\rho-1)^2 + \frac{3}{4} \psi_\mt{H} (\rho-1)^3 
\cr && 
+\frac{\psi_\mt{H}}{128} \frac{6 + 39{\psi_\mt{H}}^2 
+ 4e^{\frac{7\psi_\mt{H}}{2}} {u_\mt{H}}^2 ({\psi_\mt{H}}^2 -1)  }{{\psi_\mt{H}}^2 - 1} (\rho-1)^4 +
{\cal O} \left( {(\rho-1)}^5 \right)\,.
\label{embap}
\eea
The explicit dependence of  the dimensionless ratio $M_\mt{q}/\sqrt{\lambda} T$ for given $\psi_\mt{H}$ and $a/T$ is detailed in Fig.~\ref{plotpsi} \cite{Jahnke:2013rca} .


\section{Limiting regimes}
\label{appB}

The background geometry is known analytically in the limiting regimes of low and high temperature \cite{ALT,MT2}. This allows to obtain analytic results for the branes' embeddings in these limits, which we do next. 

\subsection{Zero temperature limit}

We start by studying the D7-branes' behavior at exactly zero temperature. The solution for the metric (in string frame) is given by eq. (2.16) of \cite{ALT}\footnote{To avoid a possible source of confusion, we have renamed $\tilde r$ the coordinate $r$ of that reference. We also renamed $w\to z$ for consistency with the rest of our paper.}
\be
ds^2 = \tilde R{^2_s} \left( \tilde r^{7/3} (-dt^2 + dx^2 + dy^2) + \tilde r^{5/3} dz^2 +
\frac{d\tilde r^2}{\tilde r^{5/3}} + \frac{12}{11} \tilde r^{1/3} d\Omega{^2_5} \right) .
\label{metricALT}
\ee
The $\rho$ coordinate is defined so that
\be
\frac{d  \tilde r^2}{\tilde r^{5/3}} + \frac{12}{11} \tilde r^{1/3} d\Omega{^2_5} \propto d  \rho^2 + \rho^2 d\Omega{^2_5} .
\ee
The condition to get this is
\be
\frac{12}{11} \frac{\tilde r^{1/3}}{\rho^2} = \frac{1}{\tilde r^{5/3}} \left( \frac{d\tilde r}{d\rho} \right)^2 ,
\ee
which leads to
\be
\tilde r = \rho^{\sqrt{\frac{12}{11}}} .
\ee
Under this change of coordinates, the metric (\ref{metricALT}) becomes
\be
\frac{ds^2}{\tilde R{^2_s}} = \rho^{\frac{14}{\sqrt{33}}} \left(-dt^2 + dx^2 + dy^2 \right) +
\rho^{\frac{10}{\sqrt{33}}} dz^2 +
\frac{12}{11} \rho^{2 \left( \frac{1}{\sqrt{33}} -1 \right)} \left(d\rho^2 + \rho^2 d\Omega{^2_5} \right) .
\ee
The induced metric on the D7-branes is obtained by replacing
\be
d\rho^2 + \rho^2 d\Omega{^2_5} \rightarrow \left( 1 + \dot R^2 \right) dr^2 + r^2 d\Omega{^2_3} ,
\ee
where, as in the main text, $\rho^2 = {R(r)}^2 + r^2$. The determinant of this metric is
\be
\sqrt{-g} = \rho^{10 \sqrt{\frac{3}{11}} -4} r^3 \sqrt{1 + \dot R^2} ,
\ee
which, together with the solution for the dilaton specified in (2.14) of \cite{ALT}
\be
\phi = \frac{2}{3} \log \tilde r + \phi_0 ,
\ee
leads to the Lagrangian
\be
\mathcal{L} \propto \left( r^2 + R^2 \right)^{\frac{n}{2}} r^3 \sqrt{1 + \dot R^2}\,,
\ee
where $n = 26 / \sqrt{33} - 4$. The equation of motion for $R(r)$ is found to be
\be
n \left( r^2 + R^2 \right)^{\frac{n-2}{2}} R r^3 \sqrt{1 + \dot R^2} =
\prt_r \left( \left( r^2 + R^2 \right)^{\frac{n}{2}} r^3 \frac{\dot R}{\sqrt{1 + \dot R^2}} \right) ,
\ee
whose solution near $r=0$ is given by
\be
R(r) = R_0 + \frac{n}{8 R_0} r^2 + {\cal O} (r^4) .
\ee
This displays a positive curvature, showing that the anisotropy has the effect of bending the branes downward even in the absence of a black hole.


\subsection{High temperature}

We now discuss the case of small values of the anisotropy parameter or, equivalently, of high temperature. In this limit, the expressions for the background fields have been previously calculated in \cite{MT2} and are given by
\bea
\phi(u) &=& -\frac{1}{4} \ln \left( 1 + u^2 \right) a^2 + {\cal O} \left(a^4 \right)\,,\cr
\cf(u)& =& \left( 1 - u^4 \right) +
\left( \frac{u^2}{3}- \left( \frac{1}{3} + \frac{5 \ln 2}{12} \right)u^4 +
\frac{1}{8}\left(1+\frac{7}{3}u^4\right) \ln \left( 1+ u^2 \right) \right)a^2 +
{\cal O} \left(a^4 \right)\,,\cr
\cb(u) &=& - \frac{1}{4} \ln \left( 1 + u^2 \right)a^2 + {\cal O} \left(a^4 \right)\,.
\eea
Using (\ref{defro2}), one can derive
\bea
u(\rho)=\frac{\sqrt{2}\rho}{\sqrt{1 + \rho^4}}
- \rho
\frac{2\rho^2 - \left( 1+\rho^4 \right) \left(1-\ln 32+10\ln (1+\rho^2) - 5 \ln(1+\rho^4) \right) }{24 \sqrt{2}{(1+\rho^4)}^{3/2}}
a^2 + {\cal O} \left(a^4 \right)\,.\cr &&
\eea
Inserting this change of coordinates into the expressions above, one can directly derive expressions for the fields to lowest order in $a$. The function $P(\rho)$ defined in (\ref{defp}) takes the form
\bea
P(\rho) &=& \frac{1}{4} \left( 1 - \frac{1}{\rho^8} \right) \cr
&&
+\frac{1-\rho^2+\rho^6-5\ln2+\rho^8(\ln32-1)-6(\rho^8-1)\ln(\rho^2+1)+3(\rho^8-1)\ln(\rho^4+1)}{48 \rho^8}a^2 \cr &&+
{\cal O} \left(a^4 \right)\,.
\eea
This is a lengthy expression and, as a consequence, the perturbation introduced by $a \neq 0$ is too complicated to be dealt with analytically. However, a near-horizon approximation gives a solution for the embedding of the form
\be
\psi(\rho) = \psi^{(0)}(\rho) + \frac{1}{32} \psi_\mt{H} \left( \rho - 1 \right)^4 a^2 +
{\cal O} \left(a^4 , {(\rho-1)}^5 \right) ,
\ee
where $\psi^{(0)}(\rho)$ is the isotropic solution.


\section{Expansion of the divergent action}
\label{series}

In this appendix, we derive the expressions (\ref{as})-(\ref{phi_det}) used in Sec.~\ref{secHR}. We start by recalling the general FG expansions of the fields \cite{reviewHR}
\begin{eqnarray}
g_{ij}&=&{g_{(0)}}_{ij}+{g_{(2)}}_{ij}v^{2}+\left({g_{(4)}}_{ij}+{h_{(4)}}_{ij}\log{v}\right)v^{4}+\mathcal{O}\left(v^{6}\right),
\cr 
\phi&=&\phi_{(0)}+\phi_{(2)}v^{2}+\left(\phi_{(4)}+\tilde{\phi}_{(4)}\log{v}\right)v^{4}+\mathcal{O}\left(v^{6}\right),
\cr
\psi&=&\psi_{(0)}v+\left(\psi_{(2)}+\tilde{\psi}_{(2)}\log{v}\right)v^3+\mathcal{O}\left(v^{5}\right).
\label{psi}
\end{eqnarray}
The axion has a constant profile in the radial direction, so that $\chi=\chi_{(0)}$.

With some algebra, it is possible to see that the coefficients of the divergences in (\ref{divergences}) are given by the following combinations of the asymptotic expansions of the fields
\bea
a_{(0)}&=&\frac{1}{4},
\cr
a_{(1)}&=&\frac{1}{4}\left({g_{(0)}}^{ij}{g_{(2)}}_{ij}+2\phi_{(2)}-2\psi_{(0)}^{2}\right),
\cr
a_{(2)}&=&-\frac{1}{2}{g_{(0)}}^{ij}{g_{(4)}}_{ij}-\frac{1}{8}({g_{(0)}}^{ij}{g_{(2)}}_{ij})^{2}+\frac{1}{4}{g_{(0)}}^{ij}{g_{(2)}}_{jk}{g_{(0)}}^{kl}{g_{(2)}}_{li}\cr && -
\frac{1}{2}\phi_{(2)}{g_{(0)}}^{ij}{g_{(2)}}_{ij}-\frac{1}{2}\phi_{(2)}^{2}+\frac{1}{2}\psi_{(0)}^{2}{g_{(0)}}^{ij}{g_{(2)}}_{ij}+\phi_{(2)}\psi_{(0)}^{2}-\psi_{(0)}\tilde{\psi}_{(2)},
\cr
a_{(3)}&=&-\frac{1}{4}\left({g_{(0)}}^{ij}{h_{(4)}}_{ij}+2\tilde{\phi}_{(4)}\right).
\eea
These terms in the FG expansion of the fields obey certain equations of motion given by~\cite{Yiannis}
\bea
&& {g_{(2)}}_{ij}=\frac{e^{2\phi_{(0)}}}{4}\left(\partial_{i}\chi_{(0)}\partial_{j}\chi_{(0)}-\frac{1}{6}{g_{(0)}}_{ij}{g_{(0)}}^{kl}\partial_{k}\chi_{(0)}\partial_{l}\chi_{(0)}\right),
\cr
&& {h_{(4)}}_{ij}=\frac{e^{4\phi_{(0)}}}{6}{g_{(0)}}^{kl}\partial_{k}\chi_{(0)}\partial_{l}\chi_{(0)}
\left(\partial_{i}\chi_{(0)}\partial_{j}\chi_{(0)}-\frac{1}{4}{g_{(0)}}_{ij}{g_{(0)}}^{ab}\partial_{a}\chi_{(0)}\partial_{b}\chi_{(0)}\right),
\cr
&& {g_{(0)}}^{ij}{g_{(4)}}_{ij}=-\frac{11e^{4\phi_{(0)}}}{576}({g_{(0)}}^{ij}\partial_{i}\chi_{(0)}\partial_{j}\chi_{(0)})^{2},
\label{g_4}
\cr
&& \phi_{(2)}=-\frac{e^{2\phi_{(0)}}}{4}{g_{(0)}}^{ij}\partial_{i}\chi_{(0)}\partial_{j}\chi_{(0)},
 \label{phi_2}
\cr
&& \tilde{\phi}_{(4)}=-\frac{e^{4\phi_{(0)}}}{6}({g_{(0)}}^{ij}\partial_{i}\chi_{(0)}\partial_{j}\chi_{(0)})^{2},
\label{phi_tilde_4}
\cr
&& \tilde{\psi}_{(2)}=-\psi_{(0)}\left(\phi_{(2)}+\frac{1}{2}{g_{(0)}}^{ij}{g_{(2)}}_{ij}\right).
\label{psi_tilde_2}
\eea
Moreover, one can check that
\bea
&&{g_{(0)}}^{ij}{g_{(2)}}_{ij}=\frac{e^{2\phi_{(0)}}}{12}{g_{(0)}}^{ij}\partial_{i}\chi_{(0)}\partial_{j}\chi_{(0)}\,,
\qquad  {g_{(0)}}^{ij}{h_{(4)}}_{ij}=0\,,
\cr
&& {g_{(0)}}^{ij}{g_{(2)}}_{jk}{g_{(0)}}^{kl}{g_{(2)}}_{li}=\frac{7e^{4\phi_{(0)}}}{144}({g_{(0)}}^{ij}\partial_{i}\chi_{(0)}\partial_{j}\chi_{(0)})^{2}\,.
\label{h_4}
\eea

These are all the ingredients needed to invert the expansions. To zero-th order in $v$ we have from \eqref{psi} that
\begin{equation}
{g_{(0)}}_{ij}=g_{ij}, \hspace{1cm} \phi_{(0)}=\phi, \hspace{1cm} \chi_{(0)}=\chi, \hspace{1cm}\psi_{(0)}=\frac{\psi}{v}.
\end{equation}
Substituting this in \eqref{g_4}-\eqref{h_4} and taking into consideration that $g_{ij}=v^{2}\gamma_{ij}$ leads to
\bea
&& {g_{(0)}}^{ij}{g_{(2)}}_{ij}=\frac{e^{2\phi}}{12v^{2}}\partial^{i}\chi\partial_{i}\chi\,,
\qquad {g_{(0)}}^{ij}{g_{(4)}}_{ij}=-\frac{11e^{4\phi}}{576v^{4}}(\partial^{i}\chi\partial_{i}\chi)^{2},
\cr
&&
 {g_{(0)}}^{ij}{g_{(2)}}_{jk}{g_{(0)}}^{kl}{g_{(2)}}_{li}=\frac{7e^{4\phi}}{144v^{4}}(\partial^{i}\chi\partial_{i}\chi)^{2}
\,,
\cr
&&
 \phi_{(2)}=-\frac{e^{2\phi}}{4v^{2}}\partial^{i}\chi\partial_{i}\chi\,,\qquad
\tilde{\phi}_{(4)}=-\frac{e^{4\phi}}{6v^{4}}(\partial^{i}\chi\partial_{i}\chi)^{2},
\cr
&&
\psi_{(0)}=\frac{1}{v}\psi\left(1-\frac{5e^{2\phi}}{24}\partial^{i}\chi\partial_{i}\chi\log{v}\right)\,,\qquad
\tilde{\psi}_{(2)}=\frac{5e^{2\phi}}{24v^{3}}\psi\, \partial^{i}\chi\partial_{i}\chi\,,
\label{psi_0_cuadrado_cov}
\eea
and, similarly, to 
\bea
&& \sqrt{-g_{(0)}}=v^{4}\sqrt{-\gamma}\left(1-\frac{e^{2\phi}}{24}\partial^{i}\chi\partial_{i}\chi\right),
\cr
&&
 e^{\phi_{(0)}}=e^{\phi}\left(1+\frac{e^{2\phi}}{4}\partial^{i}\chi\partial_{i}\chi+\frac{e^{4\phi}}{6}\left(\partial^{i}\chi\partial_{i}\chi\right)^{2}\log{v}\right).
\label{phi_0_cov}
\eea
Using these expressions, (\ref{as})-(\ref{phi_det}) can be obtained readily.


\end{document}